\pdfoutput=1
\documentclass[12pt,showpacs,preprintnumbers,floatfix]{article}
\usepackage{graphicx} 
\usepackage{dcolumn} 
\usepackage{bm} 
\usepackage{amsfonts}
\usepackage{amsthm}
\usepackage{amsmath}
\usepackage{amssymb}
\usepackage{epsfig}
\usepackage{color}
\usepackage{textcomp}
\usepackage{hyperref}
\usepackage{titlesec}
\usepackage{slashed}
\usepackage{caption}
\usepackage{subcaption}
\usepackage{siunitx}
\usepackage{booktabs}
\usepackage{multirow}
\usepackage{cite}
\usepackage{placeins}
\usepackage{comment}
\usepackage{braket}

\def\be{\begin{equation}}
\def\ee{\end{equation}}
\def\ba{\begin{eqnarray}}
\def\ea{\end{eqnarray}}

\def\bdm{\begin{displaymath}}
\def\edm{\end{displaymath}}

\def\bq{\begin{quote}}
\def\eq{\end{quote}}

 at 10truept

\newcommand{\bea}{\begin{eqnarray}}
\newcommand{\eea}{\end{eqnarray}}

\newcommand{\bi}{\begin{itemize}}
\newcommand{\ei}{\end{itemize}}

\newcommand{\beq}{\begin{equation}}
\newcommand{\eeq}{\end{equation}}
\newcommand{\beqa}{\begin{eqnarray}}
\newcommand{\eeqa}{\end{eqnarray}}

\def\ltap{\ \raise.3ex\hbox{$<$\kern-.75em\lower1ex\hbox{$\sim$}}\ }
\def\gtap{\ \raise.3ex\hbox{$>$\kern-.75em\lower1ex\hbox{$\sim$}}\ }
\def\gl{\ \raise.5ex\hbox{$>$}\kern-.8em\lower.5ex\hbox{$<$}\ }
\def\roughly#1{\raise.3ex\hbox{$#1$\kern-.75em\lower1ex\hbox{$\sim$}}}

\begin{document}

\vspace*{2cm}

\begin{center}
{\Large \bf Gravitational Atoms}\\

\vspace*{1.0cm} 
{Niklas G. Nielsen$^{a}$\footnote{\tt ngnielsen@cp3.sdu.dk}, Andrea Palessandro$^{a,b}$\footnote{\tt palessandro@cp3.sdu.dk}, 
Martin S. Sloth$^{a}$\footnote{\tt sloth@cp3.sdu.dk}}\\
\vspace{.5cm} {\em $^a$CP$^3$-Origins, Center for Cosmology and Particle Physics Phenomenology \\ University of Southern Denmark, Campusvej 55, 5230 Odense M, Denmark}\\
\vspace{.5cm} {\em  $^b$Rudolf Peierls Centre For Theoretical Physics, \\ University of Oxford, Parks Road, Oxford OX1 3PU, UK}

\end{center}

\begin{abstract}
\noindent 
Particles in a yet unexplored dark sector with sufficiently large mass and small gauge coupling may form purely gravitational atoms (quantum gravitational  bound states) with a rich phenomenology. In particular, we investigate the possibility of having an observable signal of gravitational waves or ultra high energy cosmic rays from the decay of gravitational atoms. We show that if ordinary Einstein gravity holds up to the Planck scale, then, within the $\Lambda \text{CDM}$ model, the frequency of the gravitational wave signal produced by the decays is always higher than $10^{13} \, \text{Hz}$. An observable signal of gravitational waves with smaller frequency from such decays, in addition to probing near Planckian dark physics, would also imply a departure from Einstein gravity near the Planck scale or an early epoch of non-standard cosmology. As an example, we consider an early universe cosmology with a matter-dominated phase, violating our assumption that the universe is radiation dominated after reheating, which gives a signal in an interesting frequency range for near Planckian bound states. We also show how gravitational atoms arise in the minimal PIDM scenario and compute their gravitational wave signature.
\end{abstract}

\section{Introduction} 
Our universe contains organised structures on a vast range of scales, from small systems like planets and stars, to galaxies that contains trillions of stars, all the way up to extremely large structures like superclusters that encompass hundreds of thousands of galaxies. These structures are large gravitationally bound systems whose behaviour can be described classically. In this work we will entertain the idea of \textit{quantum} gravitational bound states (atoms) of \textit{elementary} particles, and how their existence can be tested experimentally. A precursor to this idea can be found in the ``gravitational atom" of \cite{Arvanitaki:2014wva}, with a superradiant black hole nucleus surrounded by an axion cloud.

Our motivation for considering gravitational atoms comes from the Planckian Interacting Dark Matter (PIDM) scenario \cite{Garny:2015sjg,Garny:2017kha,Garny:2018grs} (for related subsequent work see \cite{PIDMrel}). In the minimal PIDM model a GUT-scale scalar particle with only gravitational interactions can be produced by thermal scattering in the early universe plasma and with the right abundance to make up all of the dark matter today. As we will see, the same mechanism will also produce gravitational atoms, which will quickly decay to gravitational waves with a well defined frequency and amplitude. These gravitational waves will typically have a very high frequency compared to what can be probed with present-day techniques, but an intermediate matter dominated period or a non-minimal gravitational coupling of the PIDM can lower the frequency by up to 10 orders of magnitude. This opens an interesting observational window to the dark sector close to the Planck scale, allowing us to learn more about hidden physics, even if dark matter is super heavy. In the same spirit, it was also recently proposed that the PIDM could be looked for with direct detection experiments by measuring the gravitational effect of its large mass as it passes by the detector \cite{Carney:2019pza}.

The Bohr radius of a two-particle bound state held together by a central inverse square law potential, $V(r) = \alpha/r$, is
\begin{equation}\label{rB}
r_B = (\mu \alpha)^{-1},
\end{equation}
where $\mu=m_1m_2/(m_1+m_2)$ is the reduced mass of the system and $\alpha$ the coupling constant of interaction between the two particles. For the ordinary hydrogen atom, $\mu$ is just the mass of the electron and $\alpha_E=q^2/4\pi$ is the fine structure constant, giving $r_B \sim 0.53 \, \mbox{\normalfont\AA} = 5.3 \times 10^{-11} \,\text{m}$. The electrostatic and gravitational potentials in the non-relativistic limit have exactly the same form: they are both central inverse square law potentials, with the crucial difference that for gravity the coupling constant is not an independent parameter. The gravitational coupling constant depends on the mass of the two particles,
\begin{equation}
\alpha_G= \frac{m_1 m_2}{m_p^2},
\end{equation}
where $m_p$ is the Planck mass. The problem in trying to build a gravitational bound state with particles having masses close to the electron mass (or any other particle in the Standard Model) is then evident. Even if these particles interacted only gravitationally, their Bohr radius would be extremely large, going from $10^5$ times the radius of the observable universe for electrons, to a light year for the Higgs boson. No region in the universe is empty enough (or ever was) to allow this kind of bound states to live. The reason why gravitational bound states are unimportant for visible matter is that for ordinary particles $\alpha_G \ll \alpha_E$, i.e. their mass-to-charge ratio is much less than one (in Planck units): $m/q \ll m_p$. In other words, gravity is the weakest force for ordinary particles, thus microscopic bound states of these particles will always involve gauge interactions.

However, there could be heavy particles in a yet unexplored dark sector for which gravity is the strongest force, $m_X/q_X \gg m_p$, where $q_X$ is the charge of the U(1) subgroup of a generic non-abelian gauge theory in the dark sector. An example of this with $q_X=0$ and $m_X \sim 10^{-3} m_p$ is the minimal PIDM model. In this case the Bohr radius could be as small as the Hydrogen atomic radius or even smaller. For the minimal PIDM scenario for example, $r_B = 2 m_p^2/m_X^3 \sim 10^9 \, l_p \sim 10^{-26} \, m$, a truly microscopic size. Note that the existence of such a strongly gravitating particle does not constitute a violation of the Weak Gravity Conjecture, as the conjecture only requires one particle in the spectrum satisfying $m/q <m_p$: any of the particles in the visible sector will do.

Quantum gravitational bound states of these particles could in principle be created in the early universe through a variety of mechanisms. For most of this work we will not focus on the precise creation mechanism, but we will just assume an initial number density of bound states $n_{B,i}$ and explore its consequences. We will study the minimal scenario of gravitational bound states of two scalar particles that interact only gravitationally, both with themselves ($q_X=0$) and with visible matter, and are created shortly after inflation in a radiation-dominated universe which undergoes usual cosmological evolution. This minimal scenario is particularly elegant as it only has two free parameters, $m_X$ and $n_{B,i}$, which allows us to put strong constraints and make model-independent predictions. The gravitational atoms we consider in the minimal scenario are not protected by any global symmetry, therefore they are unstable and will decay to radiation after a finite lifetime. Since the mass controls both the charge and the inertia of the atoms, the lifetime depends very strongly on $m_X$, and is of order $m_p^{10}/m_X^{11}$. 

Bound states with $m_X \lesssim 10^{-6} m_p$ live much longer than the age of the universe and are thus stable on cosmic timescales. They can give rise to showers of UHE cosmic rays (and gravitons) when they decay inside the galaxy. For larger values of the mass, gravitational atoms decay early in the history of the universe and produce a gravitational wave signal which could be tested by futuristic GW detectors. We also find a universal lower bound on the mass of gravitational atoms, $m_X \gtrsim 10^{-8} m_p$. Lighter gravitational atoms cannot exist today as they would be disrupted by tidal forces in galaxies. For the minimal model, which considers atoms created in the very early universe, a different bound exists that comes from disruption by Hubble expansion: $m_X \gtrsim (H m_p^2)^{1/3}$. Depending on the energy scale of inflation, this bound can become stronger (and for the highest possible scales much stronger) than the one from tidal forces in galaxies. 

We find that the gravitational wave signal is hard to detect in the minimal scenario, as it peaks at very high frequencies, above $10^{13}$ Hz. Near Planckian atoms ($m_X \sim m_p$) decay immediately after being produced, close to reheating, and are redshifted to the present time following the standard cosmological evolution. Since the maximum reheating temperature at which they can be created is $T_{rh} \sim 10^{-3} m_p$ (the highest temperature still compatible with the non-observation of tensor modes), the frequency observed today for these atoms is $\sim 10^{13}$ Hz, which follows straightforwardly from the frequency at production redshifted from reheating to the present time, $m_p T_0/T_{rh}$. As the atoms decay immediately, the resulting signal is also strongly monochromatic. 

On the other hand, atoms with $m_X \sim 10^{-5} - 10^{-6} m_p$ decay today and they release their large rest energy in the form of non-redshifted gravitational waves and ultra high energy cosmic rays. Note that since the atoms decay today, the gravitons are not redshifted, and the frequency of the signal can be in principle as high as $10^{36}$ Hz, the frequency corresponding to $10^{-6} m_p$. The decay rate in this case is comparable to the Hubble rate, therefore the signal is more smeared out and loses its monochromaticity. In both cases the signal is located at frequencies that are far beyond what current and planned experiments are able to detect. Indeed, the GW signal produced by decaying gravitational atoms can go well beyond the frequency cutoff that is usually considered for gravitational waves. Conventional wisdom assumes a lower bound on the frequency given by $10^{-18}$ Hz, corresponding to wavelengths as large as the present Hubble radius of the universe, while the highest possible value is $10^{11}$ Hz, corresponding to the frequency of a Planck-energy graviton produced during the Planck era, and redshifted to the present time using standard cosmological evolution. The frequency cutoff for astrophysical processes is of course much lower, of order 10 kHz, so this huge frequency range is often taken as encompassing all gravitational waves that can be considered \cite{Maggiore:1999vm,Maggiore:2000gv}. In our scenario, gravitational atoms as heavy as $10^{-6} m_p$ can decay today and produce UHE gravitons, so the bound is clearly violated. Unstable massive particles can in principle also produce gravitons beyond the $10^{11}$ Hz cutoff, but, contrary to gravitational atoms, they predominantly decay to visible radiation since graviton production is planck suppressed. Gravitational atoms interact only through gravity, at least in the minimal scenario, therefore graviton production is as likely as production of any other scalar particle (decay to fermions and vectors is suppressed, as we will see). In particular, if there are no fundamental light scalars in the complete UV theory, gravitational atoms primary decay channel is to gravitons. We also don't know of any other source of istrotropic gravitational waves with a peak in the spectrum at such high frequencies.

We discuss a particular realisation of the minimal model using Planckian Interacting Dark Matter (PIDM), mentioned in the beginning. In this scenario the dark matter particle resides in a maximally decoupled sector, has no self-interactions and its mass is naturally close to the Planck scale. PIDM particles are created by freeze-in from the SM plasma at very high reheating temperatures and are always outside of thermal equilibrium. PIDM bound states are subdominantly created by the same freeze-in process. This purely gravitational production mechanism will always be present, also in more complicated scenarios, therefore one can take the number density of gravitational atoms that we compute in this model as an absolute lower limit on their abundance, if scalar particles satisfying the strong gravity condition exist in the early universe. The minimal PIDM model of gravitational atoms gives an unobservable signal, but it is nonetheless instructive as a concrete and almost model-independent scenario in which gravitational atoms can arise.

To make the signal observable for upcoming detectors, we need to modify one of the assumptions that define the minimal scenario. In the last section, we consider as an example non-standard cosmological evolution in the form of an early matter-dominated stage. This is a fairly generic prediction of string theory models of the early universe, as moduli are inevitably and abundantly produced during inflation and decay much later, reheating the visible sector and kick-starting the usual radiation phase. The intermediate matter phase can be quite long, going from reheating to Big Bang Nucleosynthesis (BBN) and spanning a huge range of scales. The universe expands faster in the matter-dominated era than it would have in the usual radiation-dominated phase, thus enhancing the redshift factor of the signal, and giving a smaller frequency. For near-Planckian atoms, the frequency falls in the range $10^7 - 10^{10}$ Hz, which could be detectable by near future experiments. We also show that a large non-minimal coupling of the PIDM to gravity brings the peak frequency down to more interesting values. In both cases, a modification of gravity or early universe cosmology is needed in order to bring the peak frequency of the signal below the $10^{13} \, \text{Hz}$ threshold.

\section{The minimal model}\label{minimalmodel}
We postulate the existence of a particle $X$ in the dark sector that satisfies $m_X/q_X \gg m_p$. This particle may form purely gravitational bound states. The simplest model of gravitational atoms that can still produce a rich phenomenology rests on three fundamental assumptions:
\begin{enumerate}
  \item $X$ has only standard gravitational interactions: all Standard Model and dark charges are equal to zero. In particular $q_X=0$ (no self-interactions) and $\xi_X=0$ (minimal coupling). Therefore the dark sector is maximally decoupled from the visible sector and $X$ particles may constitute a fraction, maybe even substantial, of cold dark matter in the universe.
  \item $X$ is a scalar particle without internal quantum numbers. Scalar field masses are unprotected against large quantum corrections, so in absence of additional new physics in the dark sector, we expect the mass $m_X$ to lie near the quantum gravity scale. 
  \item Gravitational atoms are created in the very early universe, near the end of inflation or just after, and they evolve in the usual $\Lambda$CDM cosmological model. The formation mechanism is such that only 2-particle atoms are efficiently created, and predominantly in their ground state. 
\end{enumerate}

The three assumptions define the simplest scenario in which gravitational atoms can in principle arise. This most minimal scenario entails a dark sector comprised only of $X$, a scalar particle with a mass close to the Planck scale, quantum gravity effects acting as the only UV cutoff. For simplicity we also restrict our attention to creation mechanisms which dominantly produce simple atoms, made up of only two particles. The phenomenology is independent of the spin if we only consider 2-particle bound states, therefore our assumptions that $X$ is a scalar is not overly restrictive (multi-particle states are heavily modified if the constituents are fermions due to Pauli blocking). For simplicity, we also consider a particle with no internal degrees of freedom. Atoms made up of particles with internal quantum numbers could be stable by charge conservation, changing the phenomenology. For example, a complex scalar particle may form stable as well as unstable (particle/antiparticle) bound states. This will however not affect our main conclusions, but only add an additional contraint on the initial number density of bound states from avoiding over-closing the universe today.

Assumption 3. is highly non-trivial for bosonic particles, due to Bose-Einstein condensation and the universally attractive nature of gravity. Bosons do not experience Pauli exclusion principle, which means that such an atom could always lower its energy by capturing an additional particle in its ground state. Many-particle bound states could then easily become more abundant as they are energetically favourable. In order to avoid this problem we can imagine for example that the initial number density of $X$ particles is small enough that the probability of a collision producing a multi-particle bound state is negligible. This is what happens in the PIDM model.

The theory of a two-particle gravitational bound state is a trivial modification of the usual theory of positronium. Our bound state consists of two identical particles with $m_1=m_2=m_X$ that interact only through gravity. All we have to do then is replace $\alpha_E \rightarrow \alpha_G$ and use the reduced mass $\mu=m_X/2$. The energy levels of the atom to lowest order in $\alpha_G$ are
\begin{equation}\label{energy}
E_n=-\frac{\mu \alpha_G^2}{2n^2} = -\frac{m_X}{4n^2} \left(\frac{m_X}{m_p}\right)^4,
\end{equation}
where $n$ is the principal quantum number. 

In distinction to atomic bound states, these gravitational bound states are not stable by any global symmetry, and the massive particles will annihilate into radiation. The atom can either decay to a pair of gravitons or a pair of SM particles. All decay channels are mediated by gravity. To obtain the decay rates we first need to compute the amplitudes for bound state production from SM particles and gravitons. In the non-relativistic limit and in the centre-of-mass frame, we can relate the amplitude for bound state production $\mathcal{M}^S_{BS}$ to the amplitude for the creation of free $X$ particles with opposite momenta $\mathcal{M}^S_{F}(\mathbf{k},-\mathbf{k})$:
\beq\label{BSamplitude1}
\mathcal{M}^S_{BS}=\frac{1}{\sqrt{m_X}} \int \frac{d^3k}{(2 \pi)^3} \tilde{\psi}(\mathbf{k}) \mathcal{M}^S_F(\mathbf{k},-\mathbf{k}),
\eeq
where $S$ is the spin of the incoming particles, or equivalently the spin of the decay product, and $\tilde{\psi}(\mathbf{k})$ is the momentum-space wavefunction of the ground state as a function of the conjugate 3-momentum $\mathbf{k}$. The total cross section for bound state production is 
\beq
\sigma_{BS}^S=\frac{\pi}{4 m_X^2} \sum_S | \mathcal{M}^S_{BS} |^2 \delta(s-4m_X^2),
\eeq
where the delta function enforces the constraint that the total centre-of-mass energy must equal the bound state mass $\sqrt{s} \approx 2 m_X$. 

We computed the decay rates for any spin $S$, and found that only those corresponding to $S=0$ and $S=2$ are non-zero at first order, see Appendix \ref{appendixA} for details. To lowest order in $\alpha_G$, the decay rates of a gravitational atom to scalars and gravitons are
\begin{align}\label{gamma}
\Gamma_{S}=N_0 \frac{m_X}{64} \left(\frac{m_X}{m_p}\right)^{10}=N_0 \frac{\alpha_G^5 m_X}{64}, \nonumber \\
\Gamma_G= \frac{41 m_X}{128 \pi^2} \left(\frac{m_X}{m_p}\right)^{10} = \frac{41 \alpha_G^5 m_X}{128 \pi^2},
\end{align}
where $N_0$ is the number of fundamental scalar degree of freedom in the low energy spectrum \footnote{Here low energy is defined in terms of the mass of dark sector particles $X$, therefore $N_0$ is the number of scalar particles in the visible sector with a negligible mass compared to $m_X$. $N_0=4$ for the SM, but the number could be much higher if supersymmetry and/or string theory are involved in its UV completion. At the other extreme, if the UV theory does not contain fundamental scalars, $N_0=0$ and gravitational atoms will predominantly decay to gravitons.}, the other decay channels (decay to fermions and vectors) being suppressed by an additional factor of $\alpha_G^2$. Therefore, to lowest order in $\alpha_G$, $\Gamma_{SM} \simeq \Gamma_S$ and $\Gamma_G/\Gamma_{SM} = 41/(2 \pi^2N_0) \approx 2/N_0$.  Note the very strong dependence of the decay rates on $m_X$ as a result of the ``gravitational charge" being proportional to the mass. 

It is clear that the mass of a gravitational atom cannot vary freely, as their size quickly becomes untenably large if the mass is considerably below the Planck scale. Both the size of these atoms, as encoded in the Bohr radius of (\ref{rB}), and the decay rates in (\ref{gamma}) depend only on the mass $m_X$. It is then possible to place a lower bound on the mass based on disruption of bound states due to tidal forces in galaxies. For $m_X \lesssim 10^{-6} m_p$, the lifetime $\Gamma^{-1}$ is much larger than the age of the universe and gravitational atoms are stable on cosmic timescales. They will therefore be a component of cold dark matter and participate in gravitational clustering, concentrating in the center of galaxies. In the vicinity of a massive object with mass $M$, tidal effects will disrupt bound systems with size $r_B$ when the tidal energy exceeds the binding energy, $GMm_X r_B /r^2 > m_X \alpha_G^2 /4$. For a solar mass star, bound systems are disrupted at distances smaller than 
\begin{equation}
r_d= \sqrt{\frac{4 M_\odot}{m_X^3} \left(\frac{m_p}{m_X}\right)^4}.
\end{equation}
Thus, the cross section for a collision able to split an atom is $\sigma \sim \pi r_d^2$. Tidal effects are strong enough to split most gravitational atoms in galaxies if the interaction rate $n_\odot \sigma v$ is much larger than the Hubble rate today, $t_U^{-1}$, namely if $n_\odot \sigma v t_U \gg 1$, where $n_\odot \sim 0.1 \, pc^{-3}$ is the stellar density in the galactic disk and $v \sim 300 \, \text{km}/\text{s}$ the typical velocity of virialized objects in the galaxy. This gives a lower bound on the mass, or equivalently an upper bound on the size, of gravitational atoms:
\begin{equation}
m_X \gtrsim 10^{-8} m_p, \quad  r_B \lesssim 10^{24} l_p \sim 0.1 \, \text{\AA}.
\end{equation}
It is interesting how the largest possible gravitational bound states are roughly the size of ordinary atoms. For our derivation to be consistent, we have to make sure that $r_d \gg R_\odot$, the typical radius of a star, for $m_X \lesssim 10^{-8} m_p$. This is true, as $r_d(m_X=10^{-8} m_p)\sim 10^4 R_\odot$.

The constraint above is universal and constitutes an absolute lower bound on the mass of a gravitational atom. Other model-dependent constraints are possible which may even become stronger in certain parameter ranges. For example, the tidal forces due to cosmic expansion are usually far too small to break a gravitational atom apart, but they can become relevant in the very early universe. Indeed, if gravitational atoms are created soon after inflation, they can be disrupted by rapid Hubble expansion. Specifically, a very general constraint comes from requiring that their size doesn't exceed the Hubble radius, $r_B < H^{-1}$, i.e.
\begin{equation}\label{hubblebound}
m_X \gtrsim (H m_p^2)^{1/3}.
\end{equation}
The strongest constraint on the mass comes from considering gravitational atom creation at reheating, when the Hubble rate attains its largest value. The current upper bound on the reheating scale is $H_{rh} \sim 5 \times 10^{-6} m_p$, which leads to a very strong bound on the mass, $m_X \gtrsim 0.01 \, m_p$. The bound is relaxed if one lowers the reheating scale or considers creation at a lower epoch. 
\section{Phenomenology}
We now study the experimental signatures of gravitational atoms. Immediately after reheating, in a minimal scenario  with no additional new physics, their number density will evolve according to the Boltzmann equation
\begin{align}\label{boltz}
\dot{n}_B = -3Hn_B+<\sigma v>_{SM\rightarrow B}n^2_{SM}+<\sigma v>_{X\rightarrow B}n_X^2 \nonumber \\
+<\sigma v>_{G\rightarrow B}n^2_{G}-\Gamma_{SM} n_B-\Gamma_G n_B.
\end{align}
Due to assumption 1, these bound states can only be created by gravitational scattering of SM sector particles with cross section $<\sigma v>_{SM\rightarrow B}$, dark sector particles $X$ with cross section $<\sigma v>_{X\rightarrow B}$ and gravitons with cross section $<\sigma v>_{G\rightarrow B}$. They can decay back to SM scalars and gravitons with decay rates $\Gamma_{SM}$ and $\Gamma_G$ respectively. Gravitational atoms cannot decay back to their constituent particles $X$ because of conservation of energy. 

This also means that the creation of a bound state by $X$ particles has to involve emission of external radiation either in the incoming or outgoing particles in order to conserve energy. It is one of our assumptions that $X$ interacts only gravitationally, so the emitted particle has to be a graviton, which means that $<\sigma v>_{X\rightarrow B}$ is naively suppressed compared to $<\sigma v>_{SM\rightarrow B}$ and $<\sigma v>_{G\rightarrow B}$ by a factor of $\alpha_G=(m_X/m_p)^2$. However, the emission of an external graviton opens up the phase space of the process $X \rightarrow B$, leading to an enhancement factor that could partially compensate the suppression by $\alpha_G$, as long as $m_X$ is not too small.

Here we are not interested in the precise creation mechanism, so we will limit ourselves to a couple of considerations. If gravitational atoms are created after or during reheating, they can be produced in basically two distinct regimes, depending on wether $n_{SM}$ or $n_X$ dominates the total number density of the universe. In the regime $n_{SM}\gg n_X$, the creation term $<\sigma v>_{SM\rightarrow B}n^2_{SM}$ dominates in the Boltzmann equation. The visible sector is initially in thermal equilibrium and $X$ particles are created together with gravitational bound states by freeze-in. This is the PIDM scenario that we will analyse in the last section. Conversely, if $n_X \gg n_{SM}$, gravitational atoms are not created by freeze-in, but rather by scattering of free $X$ particles in the non-equilibrium \footnote{Particles with gravitational-only interactions are never in thermal equilibrium below the Planck scale.} dark plasma through the term $<\sigma v>_{X\rightarrow B}n_X^2$ (and possibly also $<\sigma v>_{G\rightarrow B}n^2_{G}$), in analogy with what happens with ordinary atoms. We can also have an intermediate regime $n_{SM}\approx n_X$ where both effects are important. The latter two cases require additional new physics in the dark sector.

The different production channels are encoded in the cross sections of (\ref{boltz}). For our purposes, we will posit an initial number density $n_{B,i}$ of gravitational atoms and treat it as a free parameter, regardless of the precise production mechanism. If the first three terms in (\ref{boltz}) (ignoring the trivial Hubble friction term) describe bound state formation, the last two describe the part relevant for observations, namely its decay to visible matter and gravitons, which we now turn to.

\subsection{Gravitational waves ($m_X \gtrsim 10^{-6}$)}
The total decay rate of a gravitational atom is $\Gamma=\Gamma_{SM}+\Gamma_G$. After a typical lifetime $\Gamma^{-1}$ has passed, the atoms will start decaying to visible matter and gravitons, with an approximate ratio of $N_0/2$ to 1. We will consider first decay to gravitons, which produces a highly energetic gravitational wave signal. We want the signal to be detectable today, so as a very rough estimate we only consider lifetimes smaller than the age of the universe, $\Gamma^{-1} \lesssim H_0^{-1}$. This translates into a bound on the mass: $m_X \gtrsim10^{-6} m_p$. Due to the huge power of $m_X^{11}$ in the decay rate, the atoms also decay well inside the radiation-dominated phase if the mass is not extremely close to saturating the bound. Taking into account the late matter-dominated phase complicates the formulas without adding much to the discussion, therefore in the following we will consider a pure radiation-dominated universe from reheating to the present day. The plots, however, include the factor of $\sim 0.2$ which accounts for the late stage of faster matter-dominated expansion.

Atoms are created close to reheating with an initial number density $n_{B,i}$, as described by (\ref{boltz}). Assuming that the creation mechanism is fast enough, the decay process can be described separately, as it takes place after creation is over. Absorbing the expansion of the universe in the definition of the comoving number density $Y_B \equiv n_B a^3$, and neglecting the creation terms, the Boltzmann equation for decay becomes
\begin{equation}\label{boltzYB}
\frac{dY_B}{da}=-\frac{\Gamma Y_B}{a H(a)},
\end{equation}
where in the radiation dominated phase $H(a)=(T_{rh}^2/\kappa_2^2 \gamma^2 m_p) (a_{rh}/a)^2$, and we consider for simplicity instantaneous reheating with maximum efficiency $\gamma=1$. $T_{rh}$ is the reheating temperature, $\kappa_2=(45/(4 \pi^3 g_{rh}))^{1/4} \approx 0.25$, and $g_{rh}$ is the number of degrees of freedom at reheating, which we will assume to be that of the SM. The solution is
\begin{equation}\label{nB}
n_B(a)=\frac{n_{B,i}}{a^3}  \exp\left(-\frac{\kappa_2^2}{2}\frac{ \Gamma m_p}{T_{rh}^2} (a^2-1)\right).
\end{equation}
Here we normalise the scale factor at the end of reheating to 1, $a_{rh}=1$, so that $n_B(a_{rh})=n_{B,i}$. Note that in the radiation dominated phase $a\propto\sqrt{t}$, so that one retrieves the usual exponential decay law in time. Note also that since bound states are intrinsically non-relativistic objects, the condition $m_X>T_{rh}$ has to be satisfied. Since we imagine these bound states to be created in the early universe, condition (\ref{hubblebound}) is relevant and actually puts a stronger bound on the mass in the radiation dominated era, 
\beq\label{hubblebound2}
m_X \gtrsim \left(\frac{T_{rh}^2 m_p}{\kappa_2^2}\right)^{1/3}.
\eeq

\begin{figure}
\centering
\includegraphics[width=8cm]{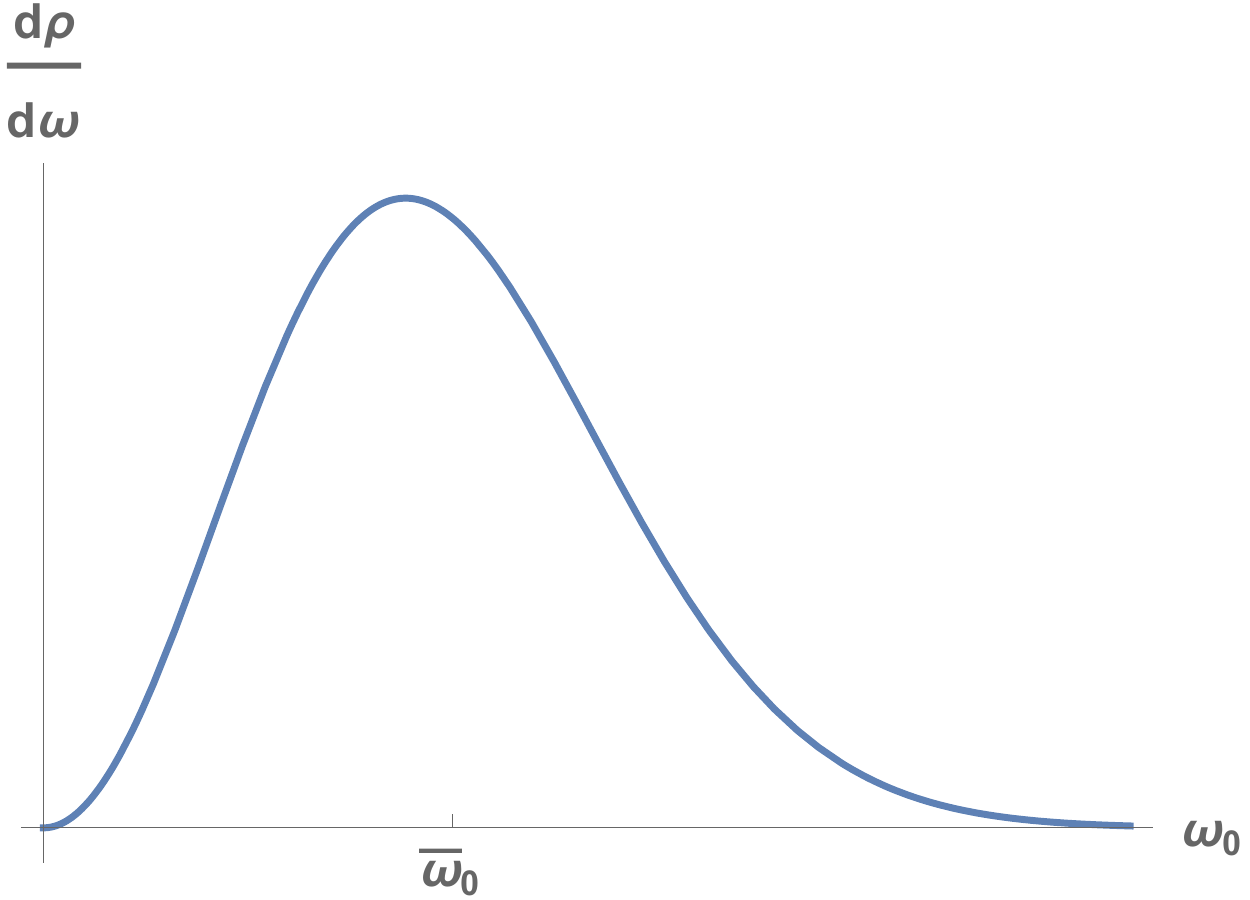}
\caption{\label{plot3} Energy spectrum of gravitational waves emitted from bound state decay in the early universe as a function of the observed frequency. The spectrum has the functional form $x^2 e^{-x^2}$, with $x=\omega_0/m_B$ and it is peaked at $\bar{\omega}_0$.}
\end{figure}

Each bound state emits gravitons with total energy equal to its mass $m_B=2 m_X$ when it decays. The infinitesimal energy density emitted by a fraction of decaying atoms is then $d \rho_G =- m_B a^{-3} dY_B$\footnote{The minus sign is necessary because while the bound state number density \textit{decreases}, the gravitational energy density \textit{increases}.}, which is then redshifted to the present value of $d \rho_{G,0} =- m_B a^{-3} dY_B (a/a_0)^4$, where $a_0=T_{rh}/T_0$ is the scale factor now, keeping in mind that $a_{rh}=1$ at the end of reheating. The redshifted frequency of a graviton emitted at a value of the scale factor $a$  today is $\omega_0=m_X (a/a_0)$. Taking into account the fact that only a fraction of energy $\Gamma_G/\Gamma$ goes into gravitons, the energy spectrum today is 
\begin{equation}
\frac{d \rho_{G,0}}{d \omega_0}=-\frac{\Gamma_G}{\Gamma}\left(\frac{a}{a_0}\right)^4 \frac{m_B}{a^3} \frac{d Y_B}{da}\frac{da}{d\omega_0}.
\end{equation}
Expressing everything in terms of $\omega_0$ we get
\begin{align}\label{spectrum}
\frac{d \rho_{G,0}}{d \omega_0}= T_0^3 \frac{\Gamma_G}{\Gamma}  \frac{n_{B,i}}{T_{rh}^3} \frac{\kappa_2^2 \Gamma m_p}{ T_0^2} \frac{\omega_0^2}{m_B^2}\exp\left[{\frac{\kappa_2^2}{2}\frac{\Gamma m_p}{T_{rh}^2}\left(1- \frac{T_{rh}^2}{ T_0^2}\frac{\omega_0^2}{m_B^2} \right)}\right].
\end{align}
The spectrum $\frac{d \rho_{G,0}}{d \omega_0}(\omega_0)$ has the form shown in Fig.\ref{plot3}.
The physical spectrum is truncated at both low and high frequencies. The minimum frequency attainable corresponds to that of a graviton emitted at production $T_{rh}$, which has a frequency $\omega_{0,\text{min}}=m_B T_0/T_{rh}$, while the maximum frequency is that of a graviton emitted today, i.e. $\omega_{0,\text{max}}=m_B$. The physical spectrum will then have these two values as a lower and upper limit. 

The spectral density is maximized at 
\begin{equation}\label{omega0*}
{\omega}^*_0=\frac{\sqrt{2}}{\kappa_2} \frac{T_0}{\sqrt{\Gamma m_p}} m_B,
\end{equation}
which corresponds to the frequency at emission redshifted by a factor of $\sim T_0/\sqrt{\Gamma m_p}$. In order for ${\omega}^*_0$ to be above the minimum frequency $\omega_{0,\text{min}}$, the condition $T_{rh} \gtrsim \sqrt{\Gamma m_p}$ has to be satisfied. If the condition is violated the maximum disappears, and the physical spectrum is just a decaying exponential peaked at $\omega_{0,\text{min}}$. We can understand the factor (\ref{omega0*}) heuristically by assuming that all atoms decay more or less at the same time $t_D \sim \Gamma^{-1}$. The overall redshift factor then is $a_d/a_0$, where $a_0=T_{rh}/T_0$ and, in the limit in which $T_{rh}\gg \sqrt{\Gamma m_p}$, $a_d \sim \sqrt{T_{rh}^2/\Gamma m_p}$, giving an approximate redshift factor of $a_d/a_0 \sim T_0/ \sqrt{\Gamma m_p}$, in accordance with (\ref{omega0*}).

In the high-mass limit, the local maximum in the spectrum disappears, and the expected frequency is well estimated by the average $\bar{\omega}_0$, 
\begin{align}\label{omega0}
\bar{\omega}_0 =(\rho_{G,0})^{-1} \int_{m_B T_0/T_{rh}}^\infty\omega_0 d \rho_{G,0}=m_B \frac{T_0}{T_{rh}} \frac{\left(1+\frac{2 T_{rh}^2}{\kappa_2^2 \Gamma} \right)}{ F\left( \frac{\kappa_2 \sqrt{\Gamma m_p}}{\sqrt{2} T_{rh}} \right)},
\end{align}
where
\begin{equation}
 F(x)= 1+ e^{x^2} \frac{\sqrt{\pi}}{2 x}\text{Erfc}\left( x\right),
\end{equation}
and the lower limit of integration is the frequency of a graviton emitted at production, as measured today.
Technically, we are only allowed to integrate up to a maximum frequency of $m_B$, which corresponds to the frequency of a graviton emitted from an atom decaying today. However, as long as $m \gtrsim 10^{-6} m_p$, we can assume that practically all bound states already decayed, and replacing $m_B$ with $+\infty$ in the integral has no effect due to the huge exponential suppression in (\ref{spectrum}). We also checked numerically that integrating to infinity has negligible effect on the final results. In the low mass-limit $T_{rh} \gg \sqrt{\Gamma m_p}$,  $\bar{\omega}_0$ agrees with (\ref{omega0*}) up to a numerical factor of order 1. In the opposite limit $T_{rh} \ll \sqrt{\Gamma m_p}$, the average frequency becomes $m_B T_0/T_{rh} \equiv \omega_{0,\text{min}}$, as atoms decay immediately after being produced, so the spectrum is a decaying exponential peaked at $ \omega_{0,\text{min}}$.

In both regimes, the spectrum is peaked at one frequency given by (\ref{omega0}), but it is not in general monochromatic, as one would expect from decays in flat space. Deviations from exact monochromaticity of the gravitational signal arise due to the stochastic nature of the decay, meaning that different atoms will decay at slightly different times and therefore be redshifted in slightly different amounts by the expansion of the universe. We can quantify the spread of the spectrum by computing the value $\delta$ at which the exponential in (\ref{spectrum}) comes to dominate,
\begin{equation}
\delta \sim \sqrt{\frac{T_0^2}{\Gamma m_p}} m_B.
\end{equation}
In the low mass regime the spectrum is fairly spread out as $\delta \sim \bar{\omega}_0$. In the high mass regime, on the other hand, $\delta / {\bar{\omega}}_0 \ll 1$, therefore the signal is strongly monochromatic and the total energy density provides a good estimate for the peak intensity of the spectrum. The total energy density in gravitational waves is just the integrated spectrum over all frequencies: 
\begin{equation}\label{rho0}
\rho_{G,0}=\int_{m_B T_0/T_{rh}}^{\infty}\frac{d \rho_{G,0}}{d \omega_0} d \omega_0=T_0^4 \frac{n_{B,i}}{T_{rh}^3} \frac{\Gamma_G}{\Gamma} \frac{m_B}{T_{rh}} F\left( \frac{\kappa_2 \sqrt{\Gamma m_p}}{\sqrt{2} T_{rh}} \right).
\end{equation}

It is clear from (\ref{omega0}) that the peak frequency today is largest for the lowest possible value of the mass, $m_X \sim 10^{-6} m_p$. The reason is that very massive bound states produce highly energetic gravitons, but they also decay very early in the history of the universe, and are therefore hugely redshifted. For example, the energy density of gravitons emitted at reheating will be redshifted by a factor $a_0^{-4}=(T_0/T_{rh})^4$, while gravitons emitted now are not redshifted at all. What is redshifted however is the number density of decaying atoms, but only by a factor $a_0^{-3}=(T_0/T_{rh})^3$. The overall enhancement factor for atoms decaying today compared to atoms decaying at reheating is thus $a_0^{-1}= T_{rh}/T_0$, which is exactly what one can see from the plots. Fig.\ref{monochromo} shows the peak intensity of the signal as the mass of the bound state increases, for atoms produced at reheating with temperature $T_{rh} \sim 10^{-3} m_p$ (saturating the experimental bound on the non-observation of tensor modes) and $T_{rh} \sim 10^{-8} m_p$. Most of the mass parameter space is excluded by condition (\ref{hubblebound2}) for $T_{rh} \sim 10^{-3} m_p$. 

\begin{figure}\label{monochromo}
\centering
\includegraphics{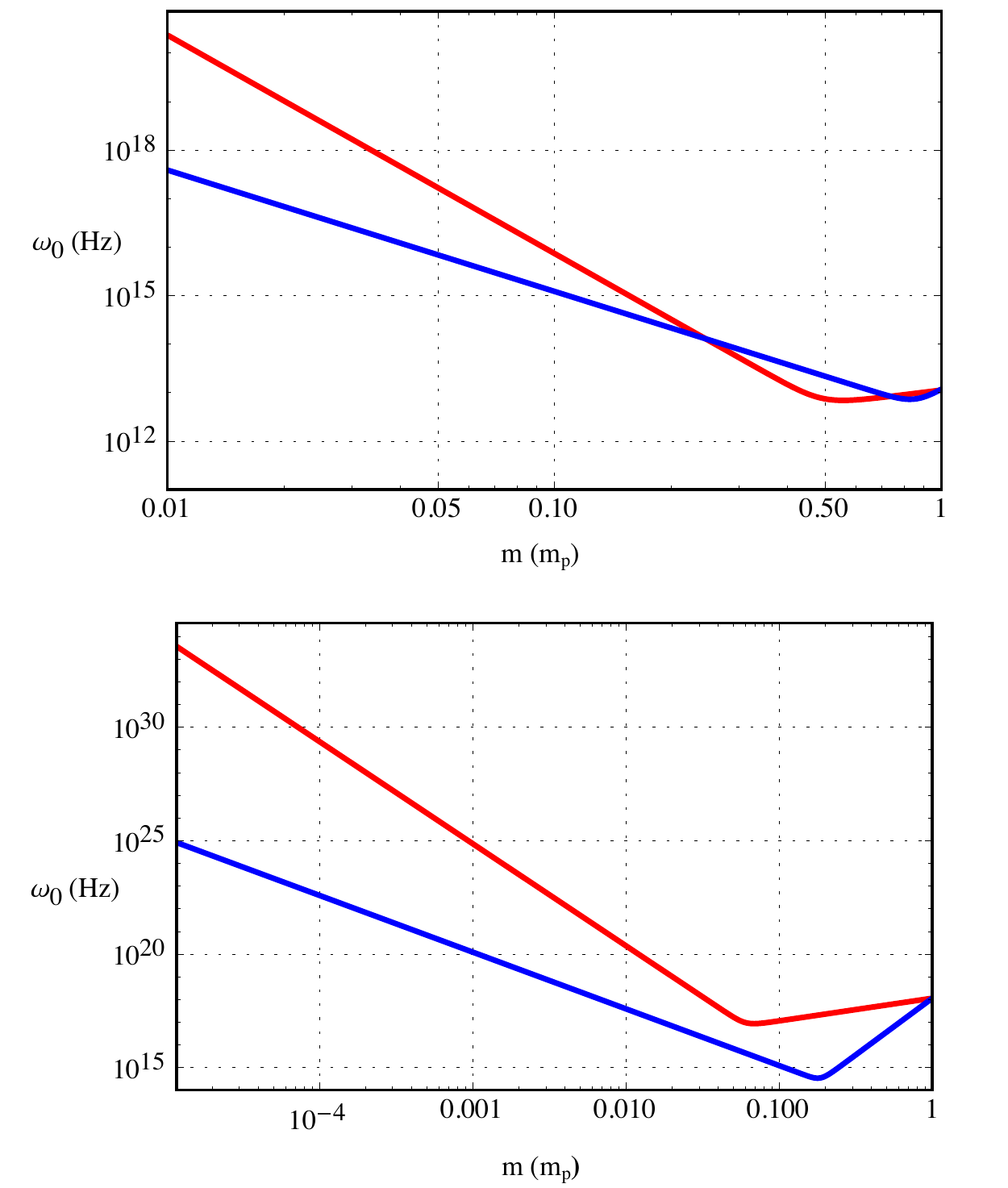}
\caption{\label{monochromo} Peak frequency of the monochromatic gravitational wave signal produced by ground state (red) and 3d excited (blue) gravitational atoms as a function of the bound state mass with $T_{rh}\sim10^{-3} \, m_p$ (top) and $T_{rh}\sim10^{-8} \, m_p$ (bottom). The frequency is in Hz and the mass in planck units. The mass range is limited by condition (\ref{hubblebound2}) coming from disruption of bound states due to Hubble expansion, $m_X \gtrsim T_{rh}^{2/3} m_p^{1/3}$. Even for $m_X$ as high as 0.5 $m_p$, the peak frequency is still at least a factor of $10^3$ larger than $T_0 \sim$ 10 GHz.}
\end{figure}

Even for atoms created at reheating with the highest possible temperature $T_{rh} \sim 10^{-3} m_p$ (so that we maximise the redshift factor) the signal is extremely energetic, well beyond the capabilities of standard interferometers like LIGO and LISA. On the other hand pilot projects are carried out with gravitational wave detectors able to observe high-frequency gravitational waves in a frequency range up to $0.1 \, \text{GHz}$ and detectors capable of measuring gravitational wave frequencies above $10^{14} \, \text{Hz}$ are also being discussed \cite{Cruise:2012zz}. The lower bound on the peak frequency of our signal for a near-planckian atom, as shown in Fig.\ref{monochromo}, is around $10^{13} \, \text{Hz}$, close to that frequency range. The reason is that, as we already discussed, the lowest frequency that we can get is for near-planckian atoms which decay to gravitons at reheating. This is just $\omega_{0,min} = m_B T_0/T_{rh} \sim m_p (T_0/T_{rh}) \gg T_0$, which is naturally much larger than $T_0 \sim 10 \, \text{GHz}$ for all allowed values of the reheating temperature. $T_0$ is actually an absolute lower bound for the frequency in the minimal model, as it corresponds to atoms being created in the Planck era and immediately decaying to gravitons which are then redshifted until today. Of course, this is already excluded by the bound on tensor modes, which places the minimum frequency at least a factor $10^3$ above the CMB temperature today. This explains the factor of $10^3$ in the figure, for $T_{rh} \sim 10^{-3} m_p$.

\begin{figure}
\centering
\includegraphics{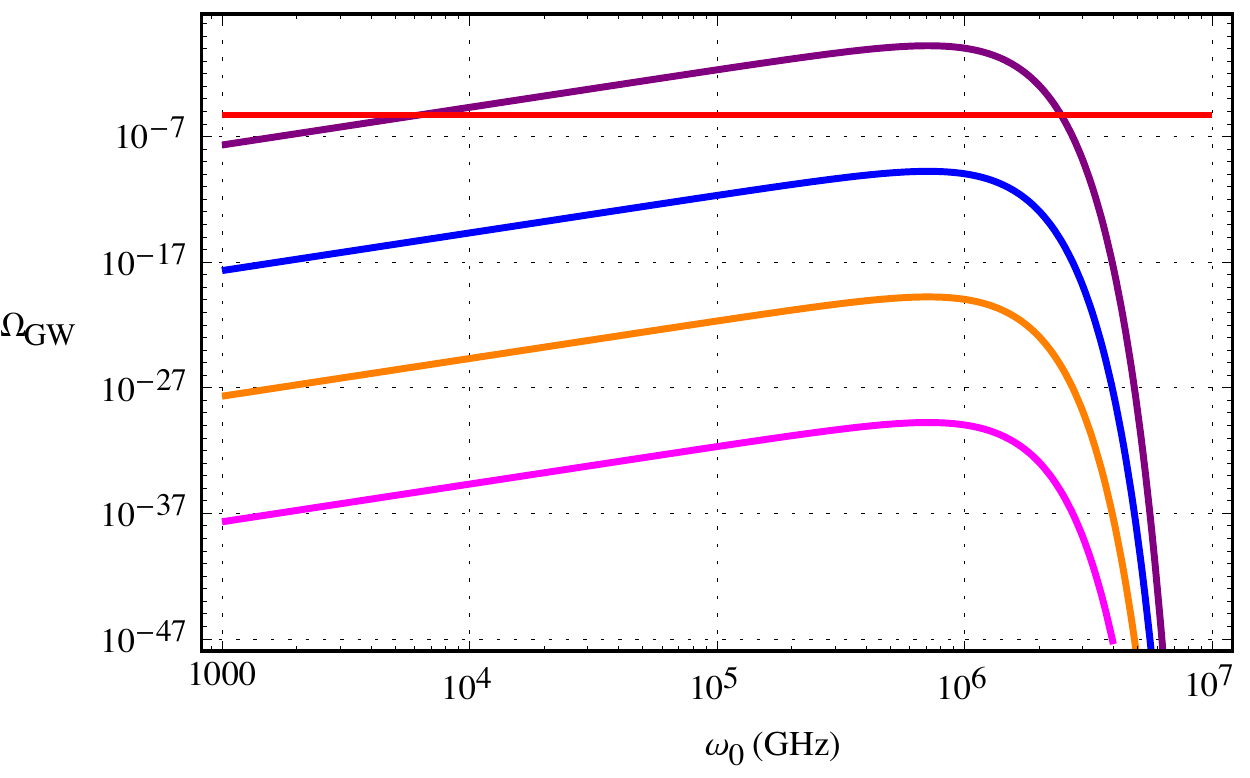}
\caption{\label{intensityplot} Gravitational wave density parameter per unit logarithmic energy $\Omega_{GW}(\omega_0)$ as a function of the signal frequency $\omega_0$, measured in units of GHz, for $T_{rh}\sim 10^{-3} m_p$, and $m_X \sim 0.1 m_p$. We parametrise the initial abundance of atoms as $n_{B,i}=\alpha T_{rh}^3$, where $\alpha$ is roughly the abundance of atoms as a fraction of the thermal plasma number density. From top to bottom we have $\alpha=1$ (purple), $\alpha=10^{-10}$ (blue), $\alpha=10^{-20}$ (orange), and $\alpha=10^{-30}$ (magenta). The red line represents the nucleosynthesis bound on the GW density parameter, $\Omega_{GW} < 5 \times 10^{-6}$.}
\end{figure}

Since we can change the number density of gravitational atoms at will, the only bound on the intensity comes from gravitational wave contribution to the effective number of neutrino species \cite{Maggiore:1999vm}. The nucleosynthesis bound, valid independently of the frequency, is $\Omega_{GW} < 5 \times 10^{-6}$, where 
\beq
\Omega_{GW}(\omega_0) = \frac{\omega_0}{\rho_c} \frac{d \rho_{G,0}}{d\omega_0}
\eeq
is the gravitational wave energy density per unit logarithmic wave frequency in units of the critical density today, $\rho_c=3 m_p^2 H_0^2$. Fig.\ref{intensityplot} shows the GW signal strength as a function of the frequency for various choices of the initial abundance of atoms $n_{B,i}$, reheating temperature $T_{rh}\sim10^{-3} m_p$ and atomic mass $m_X\sim 0.1 m_p$. We chose the values of $T_{rh}$ and $m_X$ that minimise the peak frequency. 

Naively, one could think that a way to generate gravitational waves of smaller frequencies would be to relax our assumption 3, that most bound states are created in the ground state. If one assumes that most bound states are actually created in the first (gravitationally) excited state 3d, the graviton energy released from the transition back to the ground state is of order $m_B^5/m_p^4$, i.e. $\alpha_G^2$ suppressed compared to the energy of decay. Unfortunately, this is not so simple. The transition rate from the 3d state back to the ground state can be found in \cite{Boughn:2006st,Skagerstam:2018jkw,DYSON:2013jra} and it is equal to
\begin{equation}\label{gamma3d1s}
\Gamma_{3d \rightarrow 1s} = \frac{\alpha_G^7 m_X}{2880}.
\end{equation}
A gravitational atom in the 3d energy level can also decay directly to radiation with a rate $\Gamma_{3d}$, the relevant amplitude being the one in (\ref{BSamplitude1}), with $\tilde{\psi}$ the momentum space wavefunction of the 3d state. Integrating and retaining only the lowest order term in $\alpha_G$, we find that  $\Gamma_{3d} \sim \alpha_G^4 \Gamma_{1s} \sim \alpha_G^9 m_X$. The suppression factor of $\alpha_G^4$ is due to the fact that the wavefunction of an $l=2$ state near the origin goes like $r^2$, therefore (\ref{BSamplitude1}) vanishes at first ($\alpha_G^0$) and second order ($\alpha_G^2$). See appendix \ref{appendixA} for details.

Decay of a gravitational atom in a graviton-induced excited state will then proceed through cascade decay to the ground state, which will then decay to radiation. The lifetime of an excited atom is therefore longer by a factor of $\alpha_G^{-2}$ compared to the lifetime of the atom in its ground state. If most atoms are in the 3d state, requiring that they are unstable on cosmic scales gives the new bound, $m_X \gtrsim 10^{-4} m_p$. In this mass range, we can rederive the results of this section with the trivial substitution $\Gamma \rightarrow \Gamma_{3d \rightarrow 1s}$, and noting that now the released energy in gravitons after the decay of an atom is no longer $m_B$, but,  
\begin{equation}
\Delta E=E_{3d}-E_{1s}=\frac{2}{9} \frac{m_X^5}{m_p^4}.
\end{equation}
The issue is that the minimum frequency $\omega_{0,\text{min}}=\Delta E \, (T_0/T_{rh})$ is again reached for near-planckian atoms, $m_X \sim m_p$, which means that $\omega_{0,\text{min}}$ is roughly $(T_0/T_{rh}) m_p\gg T_0$, i.e. of the same order of magnitude of the minimum frequency for ground state atoms. Moreover, while 3d excited atoms release less energy when they decay, they also decay later than ground state atoms due to the additional $\alpha_G^2$ suppression in the decay rate (\ref{gamma3d1s}), and are therefore redshifted less. The peak intensity of the signal for 3d bound states can be found, superimposed to the ground state signal, in Fig.\ref{monochromo}.

\subsection{Ultra high energy cosmic rays ($m_X \lesssim 10^{-6}$)}
If the mass of the bound states is $\gtrsim 10^{-6} m_p$, their lifetime is smaller than the age of the universe and they have all since decayed, producing a gravitational wave signal. On the other hand, if $m_X \lesssim 10^{-6} m_p$, but not much smaller, their lifetime is larger than the age of the universe, but they are still massive enough to produce ultra high energy (UHE) cosmic rays when they decay inside the galaxy \cite{Berezinsky:1997hy}. The possibility that the observed flux of UHE cosmic rays is dominantly produced by the decay of super-heavy particles has been excluded long ago based on the relative fraction of photons versus charged cosmic rays \cite{Abraham:2006ar}. Assuming instead that the observed flux is of astrophysical origin, it is possible to put stringent upper limits on a potential exotic contribution due to gravitational atoms decay.

If we define $r_X=\alpha_X t_U/\tau_X$, where $\alpha_X$ is the abundance of bound states as a fraction of the total dark matter density and $\tau_X=\Gamma^{-1}$ their lifetime, the flux of UHE cosmic rays produced by gravitational atoms decay is bounded by $r_X \lesssim 5 \cdot 10^{-11}$. If the mass is very close to $10^{-6} m_p$, then $\tau_X \sim t_U$, and the density of $X$ particles has to be rather low, of the order $\alpha_X \lesssim 10^{-12}$. If however $\alpha_X\sim1$ so that gravitational atoms constitute the dominant component of cold dark matter, then the lifetime has to be $\tau_X \gtrsim 10^{12} \, t_U \sim 10^{22} \, \text{yrs}$, which is reached for $m_X \sim 10^{-7} m_p$, only one order of magnitude below the critical value. The lowest possible mass for gravitational atoms before they start getting disrupted in galaxies is $\sim 10^{-8} m_p$, therefore the constraints above only apply to the narrow interval of masses $10^{-8} \lesssim m_X/m_p \lesssim 10^{-6}$ and for atoms in the ground state. The constraints are modified if one considers excited states, as these have much longer lifetimes, see (\ref{gamma3d1s}). For 3d excited atoms, $\tau_X \sim t_U$ is reached for $m_X \sim 10^{-4} m_p$ and $\alpha_X \lesssim 10^{-12}$, while 3d atoms making up all of the dark matter in the universe should have a mass of $m_X \lesssim 10^{-5} m_p$. This opens the possibility to GUT-scale gravitational atoms decaying today.

\section{The PIDM model}\label{PIDM}
Up until now we considered the minimal gravitational atom scenario, without committing to a particular model or creation mechanism. We will now study one realisation of the minimal scenario, in which the constituent particles of the atom are Planckian Interacting Dark Matter (PIDM). These particles, which we label $X$, are as decoupled as fundamentally allowed, having only gravitational interactions and a natural mass close to the Planck scale. PIDMs also come with a specific creation mechanism: they are produced by gravitational scattering in the thermal plasma of the Standard Model sector at the highest temperatures immediately after inflation. Gravitational atoms can be created along with free PIDMs by the same gravitational freeze-in mechanism, but with a suppressed abundance, as we describe below. The initial number density of atoms $n_{B,i}$, which we considered as a free parameter in the previous sections, is now completely fixed by the freeze-in mechanism and it only depends on the PIDM mass $m_X$ and the reheating temperature $T_{rh}$.
     
The evolution equation for the gravitational bound states, given by (\ref{boltz}), is supplemented in our model by the corresponding equation for the evolution of $X$. The set of Boltzmann equations that govern free $X$ and bound state production is 
\beq\label{boltzmanneq}
\begin{cases} 
\dot{n}_X = -3 H n_X + <\sigma v>_{SM \rightarrow X} n_{SM}^2 - <\sigma v>_{X \rightarrow SM} n_X^2 - <\sigma v>_{X \rightarrow B} n_X^2 \\
\dot{n}_B = -3Hn_B + <\sigma v>_{SM \rightarrow B} n_{SM}^2 + <\sigma v>_{X \rightarrow B} n_X^2 - \Gamma_S n_B - \Gamma_G n_B,
\end{cases}
\eeq
where free $X$ can be created from SM particles with cross section $<\sigma v>_{SM \rightarrow X}$, they can annihilate to SM particles in the time-reversed process with cross section $<\sigma v>_{X \rightarrow SM}$, and they can produce gravitational bound states with cross section $<\sigma v>_{X \rightarrow B}$. Bound states can be created either from SM particles or PIDM particles $X$ with cross sections $<\sigma v>_{SM \rightarrow B}$ and $<\sigma v>_{X \rightarrow B}$ respectively, and they can decay back to SM scalars or gravitons with decay rates $\Gamma_S$ and $\Gamma_G$ respectively. We neglected terms proportional to $n_G^2$, as gravitons are not part of the thermal bath and are thus very dilute, $n_G \ll n_{SM}$.
PIDM particles are also very dilute as they are created far outside of thermal equilibrium, $n_X \ll n_{SM}$. Bound state creation will then proceed via gravity mediated SM annihilations instead of PIDM or graviton scattering, so we can neglect the creation term $ <\sigma v>_{X \rightarrow B} n_X^2 $ in both equations. Consequently, the free PIDM number density is unaffected by bound state formation, and gravitational atoms are dominantly produced by thermal scattering of SM particles, just like free PIDMs. 

Note that our bound state formation mechanism is quite different from what is usually considered in the literature. Normally, one assumes that the dominant process for creating bound states of two particles $X$ interacting through a long-range potential is $X+X \rightarrow B + \gamma$, where $\gamma$ is the massless mediator. This is a radiative process in which the excess mass of the two particles is radiated away by a soft mediator. In our case the probability that two $X$ particles will meet to create a bound state is exceedingly small due to their suppressed number density, therefore the dominant mechanism becomes $SM+SM \rightarrow B$, i.e. creation of the bound state by gravitational annihilation of two SM (effectively) massless particles. In this case no external radiation is needed to conserve energy. See Appendix \ref{appendixA} for more details. 

Free PIDMs, like PIDM bound states, are created by gravitational freeze-in, as described in \cite{Garny:2015sjg,Garny:2017kha}. Both of them are produced non-relativistically, therefore we can work in the limit $m_X \gg T$. The cross sections for production of a scalar PIDM from SM particles are derived from the amplitudes in (\ref{free}) and in the non-relativistic limit they are given by
\begin{eqnarray}\label{sigmavS}
\langle\sigma v\rangle_{0}&=&\frac{ \pi m_X^2}{8 m_p^4}\left[\frac{3}{5}\frac{K_1^2}{K_2^2} +\frac{2}{5}+\frac{4}{5}\frac{T}{m_X}\frac{K_1}{K_2}+\frac{8}{5}\frac{T^2}{m_X^2}\right]\to \frac{\pi  m_X^2}{8 m_p^4}\,,\nonumber\\
\langle\sigma v\rangle_{1/2}=\langle\sigma v\rangle_{1}&=&\frac{4 \pi T^2}{m_p^4}\left[\frac{2}{15}\left( \frac{m_X^2}{T^2}\left( \frac{K_1^2}{K_2^2}-1\right) +3\frac{m_X}{T}\frac{K_1}{K_2}+6\right)\right]\to \frac{4 \pi  T^2}{m_p^4}\,,
\end{eqnarray}
where the modified Bessel functions are evaluated at $m_X/T$, with $T=T_{SM}$ being the temperature of the SM thermal bath and the expressions right of the arrow denote the limit for $T\ll m_X$, relevant for the massive PIDM regime. The total cross section for scalar PIDM production is
\beq
\langle \sigma v \rangle_{SM \rightarrow X} = N_0 \langle\sigma v\rangle_{0} + N_{1/2} \langle\sigma v\rangle_{1/2}+N_1\langle\sigma v\rangle_{1},
\eeq
 with $N_0$, $N_{1/2}$ and $N_1$ the number of scalar, fermion and vector degrees of freedom at the highest energies scales in our model. For the SM, $N_0=4$, $N_{1/2}=45$ and $N_1=12$. Absorbing the expansion of the universe in the definition of the comoving number density $Y_X \equiv n_X a^3 $, the final PIDM abundance $Y_X$ can be computed by integrating the Boltzmann equation (\ref{boltzmanneq}) in the approximation $n_X \ll n_{SM}$ that we discussed:
 \beq\label{YPIDM}
 Y_X= \int_1^\infty da \frac{a^2}{H(a)} \langle \sigma v \rangle_{X \rightarrow SM} (n_X^{eq})^2,
 \eeq
where we used detailed balance to replace $ \langle \sigma v \rangle_{SM \rightarrow X} \, n_{SM}^2$ by $ \langle \sigma v \rangle_{X \rightarrow SM}\, (n_X^{eq})^2$ and integrating to infinity has no effect due to the exponential suppression in $n_X^{eq}$ at low temperatures. The equilibrium density is
\beq
 n_X^{eq}=\frac{g_X}{2 \pi^2}  m_X^2 T K_2\left(\frac{m_X}{T} \right),
\eeq
and $H(a)=T_{rh}^2/(\kappa_2^2 \gamma^2 m_p) a^{-2}$. We normalise the scale factor at reheating to 1, $a_{rh}=1$, and we consider for simplicity instantaneous reheating with maximum efficiency, $\gamma=1$. In the regime $m_X \gg T$, $\langle \sigma v \rangle_{SM \rightarrow X} \sim N_0 \pi m_X^2/8 m_p^4$, $n_X^{eq} \sim (m_X T/ 2\pi)^{3/2} e^{-m_X/T} $ and the integral evaluates to 
 \beq\label{YX}
 Y_X \equiv n_{X,i} \simeq \frac{\kappa_2^2 N_0 m_X^4  T_{rh}^2}{2^7 \, m_p^3 \, \pi^2} \exp\left(-\frac{2 m_X}{T_{rh}}\right).
 \eeq
This is the initial abundance of PIDM particles after freeze-in. In the non-relativistic limit we are considering, gravitational bound states are also created with the same mechanism.

If the mass is not too close to the planck scale, the creation and decay processes are decoupled, meaning that the lifetime of gravitational atoms is much longer than the time it takes to create them from the SM bath. We can thus describe the two processes separately, as the atoms are effectively stable during creation. Taking also into account the fact that creation by $X$ scattering is completely negligible, the Boltzmann equation simplifies to
\beq\label{boltzB}
\frac{dY_B}{da} = \frac{a^2}{H(a)} <\sigma v>_{SM \rightarrow B} n_{SM}^2,
\eeq
where $n_{SM}=N_0 T^3/\pi^2$. We compute $<\sigma v>_{SM \rightarrow B}$ in appendix \ref{appendixA}, equation (\ref{gondolo}). 
We can simplify the Boltzmann equation by using the detailed balance condition in the visible sector,
\beq\label{detailed}
 <\sigma v>_{SM \rightarrow B} n_{SM}^2=\Gamma_S n_B^{eq}.
\eeq
It is a nice consistency check to ensure the validity of the condition above in the non-relativistic limit $m_X \gg T$, using the explicit formula (\ref{gondolo}). We can easily integrate the Boltzmann equation (\ref{boltzB}) in the regime $\Gamma \ll H$ to find the initial number density of gravitational atoms:
 \beq\label{YB}
 Y_B = \int_1^\infty da \frac{a^2}{H(a)} <\sigma v>_{SM \rightarrow B} n_{SM}^2= \int_1^\infty da \frac{a^2}{H(a)} \Gamma_S n_B^{eq},
 \eeq
where in the last equality we used the detailed balance condition. The formula is completely analogous to the corresponding one for the PIDM, (\ref{YPIDM}). In the non-relativistic regime we have that $\Gamma_S=N_0 m_X \alpha_G^5/64$ (the decay rate stays the same) and $n_B^{eq} \sim (m_X T/\pi)^{3/2} e^{-2m_X/T}$, so the final bound state yield is
\beq\label{YBf}
Y_B \equiv n_{B,i} \simeq \frac{\kappa_2^2 N_0 m_X^{11} \sqrt{m_X T_{rh}}}{2^7 \pi ^{3/2} m_p^9} \exp \left(-\frac{2 m_X}{T_{rh}}\right),
\eeq
and the ratio between the two is 
\beq\label{ratio}
\frac{Y_B}{Y_X} =\frac{m_X^6}{m_p^6} \sqrt{\pi \frac{m_X^3}{T_{rh}^3}}.
\eeq  
Note that the ratio is independent from $N_0$. The suppression factor of $\alpha_G^3$ comes from the corresponding suppression in the cross section for producing bound states as opposed to free particles. The bound state yield is maximised for the highest possible reheating temperature, $T_{rh} \sim 10^{-3} m_p$, and $m_X=23 \,T_{rh}/4 \approx 6 T_{rh}$ (as one can easily check by computing the derivative and putting it to zero). The best we can do therefore is to take $m_X/6 \sim T_{rh} \sim 10^{-3} m_p$. This is already contrived as the non-relativistic regime is on the verge of breaking down, but it is illustrative of the best we can hope for in this model. 

\begin{figure}
\centering
\includegraphics[width=12cm]{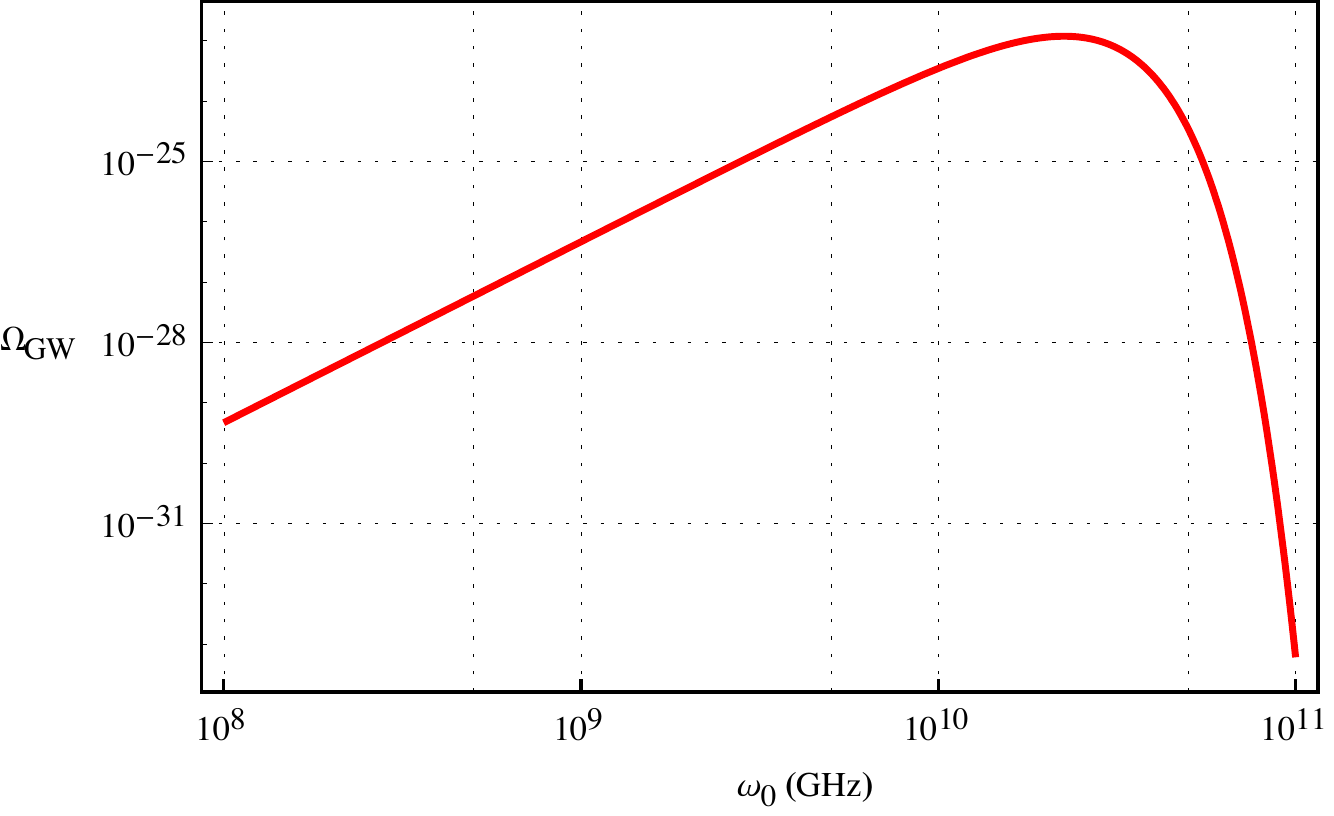}
\caption{\label{pimplot}Gravitational wave density parameter per unit logarithmic energy $\Omega_{GW}(\omega_0)$ as a function of the signal frequency $\omega_0$ (in units of GHz) in the PIDM scenario with $T_{rh}\sim 10^{-3} m_p$, and $m_X \sim 0.01 m_p$. The initial abundance of gravitational atoms is set by freeze-in, eq.(\ref{YBf}).}
\end{figure}

Now we can just take (\ref{YBf}) and plug it in the general formula for the spectrum (\ref{spectrum}). With these numbers the signal intensity is plotted in Fig.\ref{pimplot}. The intensity at the peak is in the ballpark of future experiments, but the peak frequency $\omega_0$ is unfortunately many orders of magnitude above the 10 GHz cutoff. While we are not aware of any plans to explore this frequency (see however \cite{Cruise:2012zz} for a discussion of possible experiments going beyond $10^{14} \, \text{Hz}$), the model still gives a concrete example of how gravitational atoms could arise in a sensible scenario. Moreover, freeze-in of gravitational atoms from the SM bath, being a purely gravitational process, will always be present even in more complicated scenarios for their production. Therefore one can take (\ref{YBf}) as a lower bound on their abundance if heavy scalar fields satisfying the bound $m_X/q_X > m_p$ exist. In the next section, we will consider slight modifications of the minimal scenario which give an observable signal around the 10 GHz cutoff.

\section{Modifications of the minimal scenario}
In section \ref{minimalmodel} we defined the minimal model by listing three basic assumptions, and we proved that any model that follows these assumptions will produce a gravitational wave signal at frequencies beyond the $10^{10}$ Hz threshold, the maximum frequency that will be explored by future experiments. The PIDM scenario is one instantiation of the minimal model laid out in section \ref{minimalmodel}, where the initial number density of gravitational atoms is fixed by the PIDM mass and the reheating temperature, and can be computed exactly. We now consider two extensions of the minimal model that are able to produce a signal at lower frequencies, closer to the 10 GHz mark. The first modification is universal and can be applied to any model, while the second one is specific to the PIDM scenario.

\subsection{Early matter domination}\label{matterdomination}

In this first modification, we relax assumption 3. that the universe is radiation dominated after reheating. The redshift factor for the gravitational wave signal today increases if in its early stages of evolution the universe expands faster than in the radiation dominated phase. This can be achieved for example if the early universe is dominated by non-relativistic matter from the end of inflation to Big Bang Nucleosynthesis (BBN), at which point the matter fluid decays to radiation reheating the universe at a temperature $T_{BBN} \simeq 1 \, \text{MeV}$. While BBN represents the upper limit of where we can push our matter-dominated period, the lower limit is just given by the experimental bound on the Hubble rate at inflation due to the non-observation of primordial tensor modes, i.e. $H_i \simeq 5 \times 10^{-6} m_p$. A matter-dominated period between these two scales will give the maximum enhancement to the redshift factor, if we assume that gravitational atoms are created immediately after inflation and decay very shortly after. 

An early matter-dominated phase is present, for example, in most string theory models of the early universe \cite{Angus:2013sua} and in the curvaton models \cite{Enqvist:2001zp, Lyth:2001nq, Moroi:2001ct}. A generic feature of the four-dimensional effective theories arising from compactifications of string theory is the presence of moduli, massive scalar particles with feeble, Planck suppressed interactions. Owing to their feeble Planck suppressed interactions, moduli are long-lived. They become displaced from their final metastable minimum during inflation and begin to oscillate as matter, quickly dominating the energy density. The universe then enters a modulus-dominated stage after inflation, which lasts until the moduli decay into visible matter, thus inducing reheating. The reheating temperature after thermalisation is $T_{rh} \sim \sqrt{m_{\Phi}^3/m_p}$, where $m_{\Phi}$ is the moduli mass, and reaches $T_{BBN} \sim$1 MeV for $m_\Phi \sim$ 10 TeV.

\begin{figure}
\centering
\includegraphics[width=10cm]{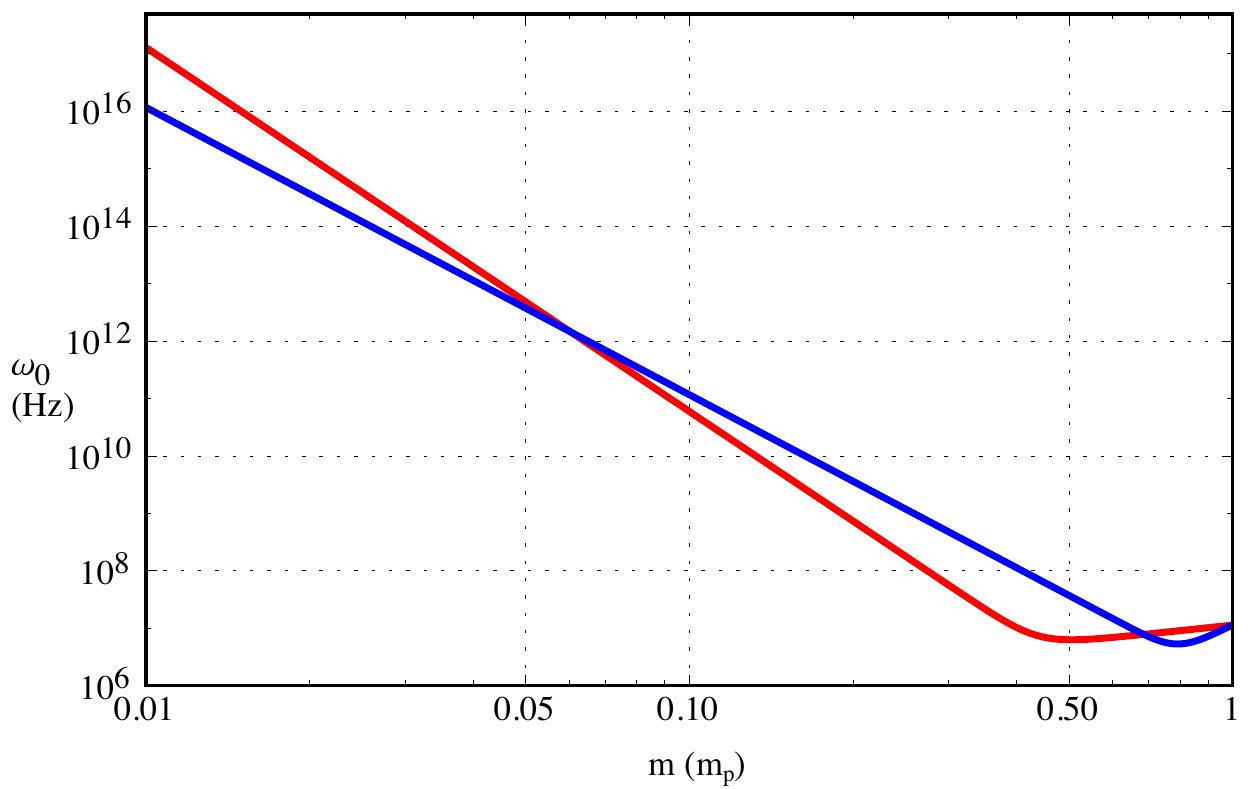}
\caption{\label{monochromo2}Peak frequency of the monochromatic gravitational wave signal produced by ground state (red) and 3d excited (blue) gravitational atoms as a function of the bound state mass with an extended period of early matter domination and $H_i\sim 5 \times 10^{-6} \, m_p$, saturating the experimental bound on tensor modes. The frequency is in Hz and the mass in planck units. The mass range is limited by condition (\ref{hubblebound2}) coming from disruption of bound states due to Hubble expansion, $m_X \gtrsim 0.01 m_p$. Due to the enhancement of the redshift factor during matter-domination, the signal frequency is pushed down to an interesting range for extremely massive atoms, $m_X \gtrsim 0.1 m_p$.}
\end{figure}

The Hubble rate in the early matter-dominated phase is $H(a)=H_i \, a^{-3/2}$, normalizing the scale factor at the end of inflation to one, $a_i=1$. The number density of atoms in this period follows from integrating (\ref{boltzYB}) with the new scale factor dependence:
\beq
n_B(a)=\frac{n_{B,i}}{a^3} \exp \left[ \frac{2 \Gamma}{3 H_i} (1-a^{3/2})\right].
\eeq
From this we can compute the spectrum in a manner that is completely analogous to what we did previously, with the difference that now the graviton is emitted during the early matter-dominated phase. The redshifted frequency of a graviton as measured today is $\omega_0 = m_B (a/a_{BBN}) (T_0/T_{BBN})$, where the scale factor at BBN $a_{BBN}$, when the universe is reheated, is related to the temperature by $a_{BBN}=(\kappa_2^2 H_i m_p/T_{BBN}^2)^{2/3}$. The spectrum is
\beq
\frac{d \rho_{G,0}}{d \omega_0} = T_0^3 \frac{\Gamma_G}{\Gamma} \frac{m_B n_{B,i}}{m_p^2 H_i^2}  \left(\frac{\omega_0}{m_B}\right)^{3/2} \frac{\Gamma m_p}{\sqrt{T_0^3 T_{BBN}}} \frac{T_{BBN}}{\kappa_2^2 m_B} \exp\left[ \frac{2 \Gamma}{3 H_i} \left( 1-\frac{H_i \kappa_2^2 m_p}{\sqrt{T_0^3 T_{BBN}}} \left(\frac{\omega_0}{m_B}\right)^{3/2} \right) \right],
\eeq
and has now the functional form $x^{3/2} e^{-x^{3/2}}$. The average frequency is
\beq
\bar{\omega}_0=m_B T_0 \frac{ E_{-\frac{4}{3}}\left(\frac{2 \Gamma }{3 H_i}\right)
  }{E_{-\frac{2}{3}}\left(\frac{2 \Gamma }{3 H_i}\right)} \sqrt[3]{\frac{T_{BBN}}{H_i^2 \kappa_2^4
   m_p^2}},
\eeq
where $E_n(x)=\int_1^\infty dt \, e^{-x t}/t^n$ is the exponential integral function. In the high mass limit $\Gamma \gg H_i$, the spectrum is peaked at $\bar{\omega}_0 \simeq m_B (T_0/T_{BBN}) a_{BBN}^{-1}$, which corresponds to the minimum frequency in this scenario (most atoms decay at $a \approx 1$). This frequency can be much smaller than (\ref{omega0}), due to the faster expansion in the matter phase. Fig.\ref{monochromo2} shows the peak frequency for both ground state and 3d excited gravitational atoms when we allow for an early matter-dominated phase immediately after inflation. As one can see, the minimum frequency drops 7 orders of magnitude, allowing us to reach an interesting frequency range, $\sim 10^7 - 10^{10} \, \text{Hz}$. While we are successful in decreasing the frequency, the price to pay for an early matter-dominated phase is a comparable drop in the energy density of the signal. The density parameter $\Omega_{GW} = \rho_{GW}/\rho_c$ now picks up a suppression factor of $a^{-1}_{BBN}=(T_{BBN}^2/\kappa_2^2 H_i m_p)^{2/3}$. 

\subsection{Non-minimally coupled PIDM}
 The minimal PIDM model for the gravitational atom is extremely constrained, having the PIDM mass $m_X$ as the only free parameter. This minimal scenario gives a gravitational wave signal in the frequency range around $\sim 10^{10}$ GHz, far too energetic to be observed by near-future detectors. Here we consider a non-minimal modification of the PIDM scenario, which decouples the interaction strength from the mass of the constituents, thereby providing a way to bring the peak frequency down to more interesting values. In particular, we drop assumption 1. that the PIDM couples minimally to gravity. 
 
 We postulate an additional non-minimal coupling to gravity, of the form
 \beq
 \mathcal{L}_{NM}=\frac{1}{2}  \xi_X X^2 R,
 \eeq
 where $X$ is the PIDM, and $\xi_X$ is the non-minimal coupling parameter. In \cite{Garny:2017kha} we computed the thermally averaged cross section in the non-relativistic limit for the production of the PIDM with a non-minimal coupling of this sort. The result is
 \beq
 <\sigma v>_{\phi \phi\rightarrow X X} \simeq \frac{G^2m_X^2\pi}{8} (1+4 \xi_X)^2 \sim 2 \pi \, G^2 m_X^2  \xi_X^2.
 \eeq
Production is enhanced by powers of the non-minimal coupling parameter. Therefore, by having a large non-minimal coupling to gravity, we can now efficiently create atoms which also decay faster due to the strong coupling. Moreover, excited atoms will also be created more abundantly, and if the non-minimal coupling is large enough, their abundances will only be mildly suppressed compared to their ground state relatives. In the following we will focus our attention on 3d excited states only.
 
In the Jordan frame, and to leading order in the Planck mass, the non-minimal coupling term gives a new dimension 5 operator in the action \cite{Hertzberg:2010dc}:
\beq
\frac{\xi_X}{m_p} X^2 \Box h,
\eeq
where $h$ is the metric perturbation. At tree-level, the amplitude for X scattering non-minimally via single graviton exchange in a particular channel scales as $\mathcal{M}_X\sim\xi_X^2 E^2/m_p^2$. As an example, the total amplitude squared for the scattering of $X$ in the s-channel is
\beq
\mathcal{M}_X^{\text{s-channel}}=\frac{4 \pi^2 G^2}{s^2} \left(2 m_X^4 + st+t^2 -2 s^2 \xi_X(1+\xi_X) -2m_X^2(s+2t+2s\xi_X)\right)^2.
\eeq
At large coupling $\xi_X \gg 1$, the amplitude becomes $\sim 16 \, G^2 \pi^2 s^2 \xi_X^4$ and we can effectively incorporate the non-minimal coupling in a redefinition of the gravitational constant $G$. Since bound state formation can be seen in quantum field theory language as summing over ladder diagrams (in t and u channels) \cite{Petraki:2015hla}, heuristically we can take into account a large non-minimal coupling of the PIDM in the non-relativistic limit by the replacement $\alpha_G \rightarrow \alpha_G \xi_X^2$ (or equivalently $G \rightarrow G \xi_X^2$). We thus redefine the gravitational constant as $\tilde{\alpha}_G \equiv \alpha_G \xi_X^2$ and use this as the new effective coupling, keeping in mind that the unitarity bound now reads $\tilde{\alpha}_G=\alpha_G \xi_X^2<1$. We have to be careful about blindly renormalising the gravitational constant in this way though. If the process we are evaluating involves creation or decay of the PIDM by/to other minimally coupled particles, we should remember to multiply the final amplitude squared by a factor of $\xi_X^{-2}$, since these particles only see the true bare value of the gravitational constant and they don't contribute to the enhancement.

The abundances of free PIDMs and gravitational atoms in their ground state are given by equations (\ref{YX}) and (\ref{YBf}), rescaled by a factor of $\xi_X^2$ and $\xi_X^8$ respectively. We can compute the abundance of 3d excited atoms in a similar way, using equation (\ref{YB}) with a different decay rate $\Gamma_{3d}\sim \alpha_G^9 m_X$ (we compute this exactly in the appendix). The result is just the ground state number density rescaled by a global factor of $\xi_X^{-2} \tilde{\alpha}_G^4/(2^3 3^9 \pi)$, which accounts for the difference in the decay rates. If the value of $\xi_X$ is sufficiently large, the effective gravitational coupling can be very close to 1, and the production of ground state and first excited state gravitational atoms will be only mildly suppressed. Moreover, the decay rates will also be enhanced by the new renormalised gravitational strength, so the atoms will decay faster. 

\begin{figure}
\centering
\includegraphics[width=12cm]{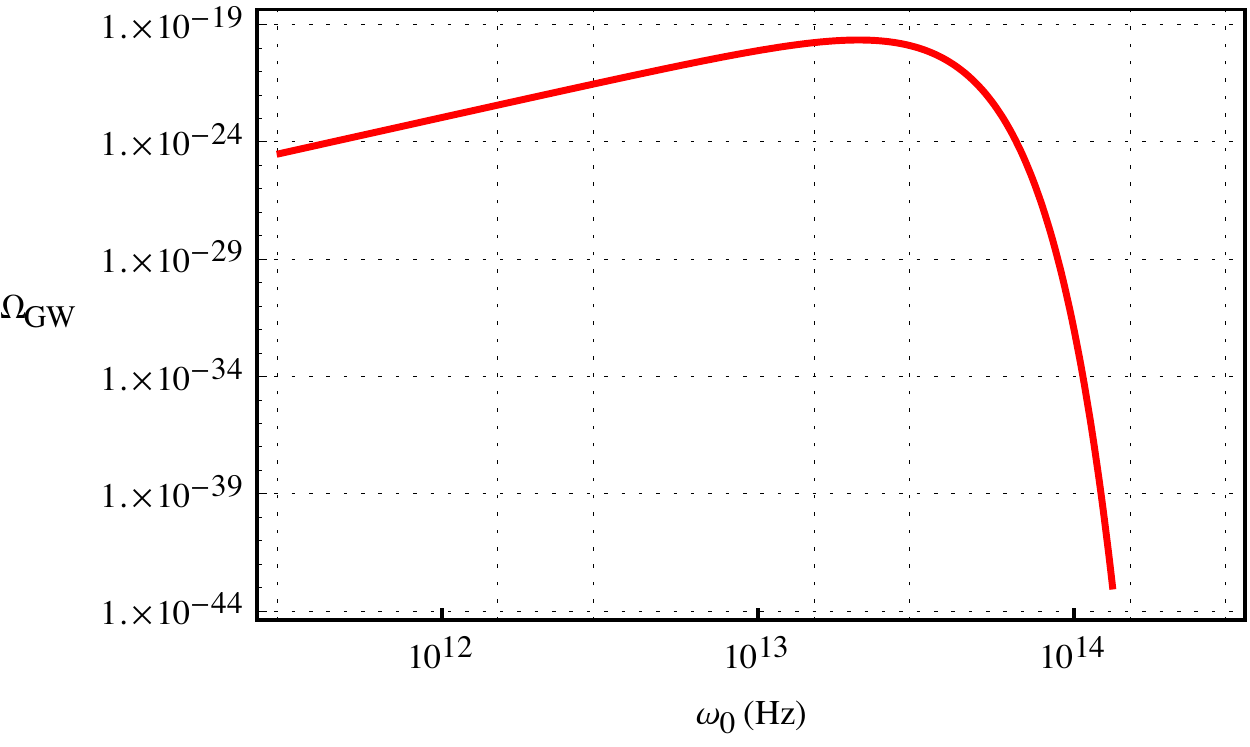}
\caption{\label{nonminPIDM}Gravitational wave density parameter per unit logarithmic energy $\Omega_{GW}(\omega_0)$ as a function of the signal frequency $\omega_0$ (in units of Hz) in the non-minimally coupled PIDM scenario with $T_{rh}\sim 10^{-3} m_p$, $m_X \sim 0.01 m_p$ and $\xi_X\sim100$. The spectrum is truncated at 1000 GHz, since this is the lowest possible frequency attainable in this scenario, corresponding to atoms decaying immediately after being produced.}
\end{figure}

Fig.\ref{nonminPIDM} shows the total intensity of the GW signal for PIDM atoms (ground state + 3d excited) with mass $m_X \sim 0.01 m_p$, reheating temperature $T_{rh}\sim10^{-3} m_p$ and non-minimal coupling $\xi_X\lesssim100$. The effective gravitational coupling is $\tilde{\alpha}_G\lesssim1$. 3d atoms relax to the ground state with a decay rate $\Gamma_{3d\rightarrow1s} \sim \xi_X^{-2} \tilde{\alpha}_G^7 m_X$ (see eq.(\ref{gamma3d1s})), releasing gravitons with energy $\sim \tilde{\alpha}_G^2 m_X$. The minimum frequency is now of order $10^{13}$ Hz, and the signal intensity drops sharply after that point. We see that a large non-minimal coupling allows us to bring the peak frequency down 10 orders of magnitude, in a range that will be explored by planned GW experiments.

\section{Conclusions}
In this work we have studied the gravitational wave signature left behind by purely gravitational atoms decaying in the very early universe. We focused on the minimal scenario in which the particles making up the atom are scalars and are only gravitationally interacting. Near-planckian atoms decay to gravitons immediately after being produced, creating a nearly monochromatic, isotropic and highly energetic gravitational wave signal. If Einstein gravity is valid all the way up to the Planck scale, and the gravitational waves are redshifted from the earliest moments after inflation until today using the standard $\Lambda$CDM scenario, the minimum frequency attainable in this scenario is $10^{13} \, \text{Hz}$, three orders of magnitude above the expected cutoff from primordial gravitational waves. This constitutes a unique source of istrotropic gravitational waves with a peak in the spectrum at such high frequencies (see \cite{Cruise:2012zz} for a discussion on how to reach this futuristic frequency sensitivity). We study in detail the minimal PIDM scenario for gravitational atom production, which gives a definite prediction for both the frequency and the amplitude of the signal. If these gravitational waves are observed at frequencies below $10^{13} \, \text{Hz}$, it would imply a non-standard early cosmological evolution or modified gravity near the Planck scale, and it would therefore give us clues about near planckian dark physics. As an example, we consider in the text an early matter dominated period and a large non-minimal coupling for the PIDM. Both break the assumptions of the minimal model, and are concrete examples of non-standard physics, which lead to lower frequencies.

\subsection*{Acknowledgements}
We would like to thank Joseph Conlon, Mathias Garny, Nemanja Kaloper, Florian Niedermann and McCullen Sandora for helpful comments. AP also wishes to thank Joseph Conlon and the Oxford particle theory group for the kind hospitality while part of this work was completed. 
This work is supported by Villum Fonden grant 13384. CP3-Origins is partially funded by the Danish National Research Foundation, grant number DNRF90. 

\newpage
\begin{appendix}
\numberwithin{equation}{section}
\setcounter{equation}{0}

\section{Amplitudes and decay rates computation}\label{appendixA}
In this appendix we compute the decay rates for a gravitational atom decaying to gravitons and SM particles. In the minimal scenario, the atom is a bound state of two scalar particles $X$ with mass $m_X$. The amplitudes squared for production of a scalar $X$ from (exactly massless) scalar, fermion and vector SM particles are, respectively \cite{Garny:2017kha}

\begin{equation}\label{free}
\begin{aligned}
  |\mathcal{M}_{0 \rightarrow 0}|^2=4G^2\pi^2\frac{(m_X^2-t)^2(m_X^2-s-t)^2}{s^2},\\
  |\mathcal{M}_{1/2 \rightarrow 0}|^2=-8G^2\pi^2\frac{(2m_X^2-s-2t)^2(m_X^4-2m_X^2t+t(s+t))}{s^2},\\
 |\mathcal{M}_{1 \rightarrow 0}|^2=8G^2\pi^2\frac{(m_X^4-2m_X^2t+t(s+t))^2}{s^2},
\end{aligned}
\end{equation}

where $s=(p_1+p_2)^2=(k_1+k_2)^2=E_{CM}^2$ and $t=(p_1-k_1)^2=(p_2-k_2)^2$ are the Mandelstam variables, $p_1$,$p_2$ are the 4-momenta of the incoming SM particles and $k_1$,$k_2$ the 4-momenta of the outgoing $X$ particles. Newton's constant in natural units is just $G=m_p^{-2}$.

We first compute the cross section for producing a $XX$ bound state directly from SM particles annihilation. From that, we will easily obtain the decay rate by going to the time-reversed process. Since SM particles are basically massless compared to $X$, bound state formation can happen without emission of external radiation. Schematically, we can write the amplitudes for producing free $X$ particles as $\mathcal{M}^S_{F}=\braket{SM(\mathbf{p_1},S)SM(\mathbf{p_2},S)|X(\mathbf{k_1})X(\mathbf{k_2})}$, where the superscript $S=0,1/2,1$ denotes the spin of the SM particles. Squaring these amplitudes gives the results in (\ref{free}). The goal now is to write the bound state in terms of free-particle states, so that the final bound state formation amplitude will just be a sum over single-particle production amplitudes. 

For a two-body system with equal masses, the center-of-mass and relative coordinates are
\beq
\mathbf{R}=\frac{1}{2} (\mathbf{r_1}+\mathbf{r_2}), \quad \mathbf{r}=\mathbf{r_1}-\mathbf{r_2},
\eeq

with conjugate momenta
\beq
\mathbf{K}= \mathbf{k_1}+\mathbf{k_2}, \quad \mathbf{k}=\frac{1}{2}(\mathbf{k_1}-\mathbf{k_2}).
\eeq

In the center-of-mass frame the total momentum $\mathbf{K}$ is zero, so $\mathbf{k_2}=-\mathbf{k_1}$, and $\mathbf{k_1} \equiv \mathbf{k}$. For a non-relativistic bound state $|\mathbf{k}| \ll m$ and $s=E_{CM}^2\approx4m_X^2$. In this regime, we can write a generic bound state with mass $2m_X$ and total momentum $\mathbf{K}=0$ as a superposition of free two-particle states with opposite momenta \cite{Peskin:1995ev}
\beq\label{BSket}
\ket{B}=\frac{1}{\sqrt{m_X}} \int \frac{d^3k}{(2 \pi)^3} \widetilde{\psi}(\mathbf{k}) \ket{\mathbf{k},\mathbf{-k}},
\eeq
where $\ket{\mathbf{k},\mathbf{-k}}$ are the free particle states and $\widetilde{\psi}(\mathbf{k})$ is the Fourier transform of the position-space Schrodinger wavefunction for the bound state:
\beq
\widetilde{\psi}(\mathbf{k})=\int d^3 x e^{i \mathbf{k} \cdot \mathbf{r}} \psi(\mathbf{r}).
\eeq
The amplitude for the bound state production is $\mathcal{M}^S_{BS}=\braket{SM(p_1,S)SM(p_2,S)|B}$, which, using (\ref{BSket}), is just
\beq\label{BSamplitude}
\mathcal{M}^S_{BS}=\frac{1}{\sqrt{m_X}} \int \frac{d^3k}{(2 \pi)^3} \widetilde{\psi}(\mathbf{k}) \mathcal{M}_{F}^S(\mathbf{k},\mathbf{-k}).
\eeq

Therefore we can just take the free amplitudes in (\ref{free}) and plug them in (\ref{BSamplitude}) with the replacement $-\mathbf{k_2}=\mathbf{k_1}\equiv \mathbf{k}$. In principle one should integrate the free amplitude over the conjugate momentum, weighted by the bound state wavefunction in Fourier space $\widetilde{\psi}(\mathbf{k})$. In practice the calculation is made much easier by noting that for a non-relativistic bound state the energy is dominated by the mass term, so that $\mathcal{M}_{F}$ roughly coincides with the amplitude for producing the $X$ particles at rest, $\mathcal{M}_{F}^S(\mathbf{k},\mathbf{-k}) \approx \mathcal{M}_{F}^S(\mathbf{0},\mathbf{0})$. In other words, $\mathcal{M}_{F}^S$ is basically constant over the integration region where $\widetilde{\psi}(\mathbf{k})$ is appreciably non-zero and we can take it out of the integral. Indeed, the typical momentum of a particle in a gravitational bound state is $|\mathbf{k}|_B=\alpha_G m \sim m_X^3/m_p^2$, which is clearly negligible compared to the mass term. Then, the integral over $\mathbf{k}$ just gives $\psi_0(0)$, the position-space wavefunction of the ground state evaluated at the origin, and (\ref{BSamplitude}) becomes

\beq\label{MBS}
\mathcal{M}_{BS}^S=\frac{1}{\sqrt{m_X}} \mathcal{M}^{S}_{F}(\mathbf{0},\mathbf{0}) \psi_0(0).
\eeq

The final expression in this case is extremely simple: the amplitude for the creation of a non-relativistic $X$ bound state from SM particles of spin $S$ is proportional to the amplitude for the creation of free $X$ at rest, the constant of proportionality being $\psi_0(0)/\sqrt{m_X}$. Clearly, the total amplitude squared, averaged over spin states, is $|\mathcal{M}_{BS}^S|^2=|\mathcal{M}^{S}_{F}(\mathbf{0},\mathbf{0})|^2 |\psi_0(0)|^2/m_X.$ Plugging this into the expression for the total cross section we obtain \cite{Peskin:1995ev}:
\beq\label{sigmaBS}
\sigma_{BS}^S=\frac{1}{8m_X^2} \int \frac{d^3 K}{(2 \pi)^3} \frac{1}{4m_X} (2 \pi)^4 \delta^{(4)}(p_1+p_2-K) |\mathcal{M}_{BS}^S|^2.
\eeq

The phase space integral removes only three of the four delta functions and, upon rewriting the last delta function using $\delta(P^0-K^0)=2K^0\delta(P^2-K^2)$, we are left with
\beq\label{sigmaBS2}
\sigma_{BS}^S= \frac{\pi}{4 m_X^2} |\mathcal{M}_{BS}^S|^2 \delta(s-4m_X^2)=\frac{\pi}{4 m_X^3} |\mathcal{M}^{S}_{F}(\mathbf{0},\mathbf{0})|^2 |\psi_0(0)|^2 \delta(s-4m_X^2).
\eeq
The last delta function enforces the constraint that the total center-of-mass energy must equal the bound-state mass $M \approx 2m_X$. We can now compute these cross sections explicitly for every value of $S=0,1/2,1$. The amplitudes squared $|\mathcal{M}^{S}_{F}(\mathbf{0},\mathbf{0})|^2$ are just the ones in (\ref{free}) for $\mathbf{k}=0$, i.e. for $s=4m_X^2$ and $t=-m_X^2$. A quick calculation gives \footnote{From now on we drop the explicit dependence of the amplitude on $\mathbf{k}$. It is understood that $\mathbf{k}=0$.}  $|\mathcal{M}^{0}_{F}|^2=4G^2\pi^2m_X^4$, $|\mathcal{M}^{1/2}_{F}|^2=0$ and $|\mathcal{M}^{1}_{F}|^2=0$. This tells us that the formation of a non-relativistic scalar $X$ bound state by two SM particles is only efficient when the SM particles are scalars: bound state creation by SM fermions and vectors is suppressed.

To get an order of magnitude estimate of the suppression, we imagine $\widetilde{\psi}(\mathbf{k})$ to be sharply peaked at $\mathbf{k}_B$ in (\ref{BSamplitude}), with $|\mathbf{k}|_B=\alpha_G m_X$. Then after integrating over $k$ we get $\sqrt{m_X} \mathcal{M}_{BS}^S \sim \mathcal{M}_F^S(\mathbf{k}_B,\mathbf{-k}_B)$, which we can now expand around $\mathbf{k}_B=0$. For $S=1/2$ and $S=1$ the zeroth-order term $\mathcal{M}_F^S(\mathbf{0})$ vanishes, so we get $\mathcal{M}_F^S(\mathbf{k}_B) \approx (1/2) \partial_{k_B^2} \mathcal{M}_F^S(\mathbf{0}) k_B^2$, where $\partial_{k_B^2} \mathcal{M}_F^S(\mathbf{0})$ is the derivative of the amplitude with respect to $k_B^2$ evaluated at $\mathbf{k}_B=0$. The amplitudes are dimensionless and only contain the mass $m_X$ as a dimensionful parameter, so by dimensional analysis $\partial_{k_B^2} \mathcal{M}_F^S(\mathbf{0})$ has to be proportional to $m_X^{-2}$, which means that the first non-zero term in the expansion for fermion and vector SM particles is of order $\sim k_B^2/m_X^2 = \alpha_G^2$, or $\alpha_G^4$ for the amplitude squared. Comparing this to the amplitude squared for scalar SM, $|\mathcal{M}^{0}_{F}|^2=4G^2\pi^2m_X^4 \sim \alpha_G^2$, the suppression factor is of order $\alpha_G^2 \sim m_X^4/m_p^4$, which is already $10^{-12}$ for a GUT scale $X$.

We could also try to compute the bound state amplitude in (\ref{BSamplitude}) without doing any approximation. For that, we need to evaluate the bound state wavefunction in Fourier space, which is easily done by solving the non-relativistic Schrodinger equation with a gravitational potential. This has the exact same form as the electrostatic potential in the non-relativistic limit, so that we can simply borrow the result from the hydrogen atom case, with the trivial replacement $\alpha_{EM} \rightarrow \alpha_G=m_X^2/m_p^2$ and bearing in mind that in our case the reduced mass of the system is $\mu=m_X/2$:
\beq
\widetilde{\psi}(\mathbf{k})=\frac{8 \sqrt{\pi } \alpha_G ^4 \mu ^4}{(\alpha_G  \mu )^{3/2} \left(\alpha_G ^2 \mu^2+k^2\right)^2}.
\eeq
We also need the free amplitude $\mathcal{M}_{F}^S(\mathbf{k},\mathbf{-k})$ as a function of the conjugate momentum $\mathbf{k}$. Let's consider the scalar case $S=0$. In this case $\mathcal{M}_{F}^0(\mathbf{k},\mathbf{-k})=-i \pi  G \left(k^2 \cos (2 \theta )-k^2-2 m_X^2\right)$. The problem is that by a simple power counting argument, the integral in (\ref{BSamplitude}) linearly diverges. We can still evaluate the amplitude in closed form by placing a hard cutoff $\Lambda$ in the integral:
\beq\label{M0BS}
\begin{split}
\mathcal{M}_{BS}^0=\frac{i G \sqrt{m_X} (\alpha_G  m_X)^{3/2}}{3\sqrt{2 \pi } \left(4 \Lambda ^2+\alpha_G ^2 m_X^2\right)} [2 \alpha_G  \Lambda  \left(8 \Lambda ^2+3\left(\alpha_G^2-2\right) m_X^2\right) \\
 -3 \left(\alpha_G ^2-2\right) m_X \left(4\Lambda^2+\alpha_G ^2 m_X^2\right) \tan ^{-1}\left(\frac{2 \Lambda }{\alpha_G  m_X}\right)].
\end{split}
\eeq
It is clear now from the explicit form of the cut-off amplitude that it diverges as $\propto \Lambda$. Moreover, it is not hard to see that the integral starts linearly diverging when $\Lambda$ approaches $m_X$. The reason for this is that we are using a non-relativistic formula for the bound state amplitude that is valid only up to energies comparable to the mass of the bound state. Beyond that, we enter the relativistic regime and our formula breaks down, giving nonsensical results. Physically, we expect an exponential suppression in $k$ to appear in the formula for the relativistic bound state wavefunction for momenta greater than $m_X$. This quickly kills the integrand function and gives a final result that is numerically not too different from the classical one cut-off at $\Lambda \sim m_X$. 

\begin{figure}
\centering
\includegraphics[width=10cm]{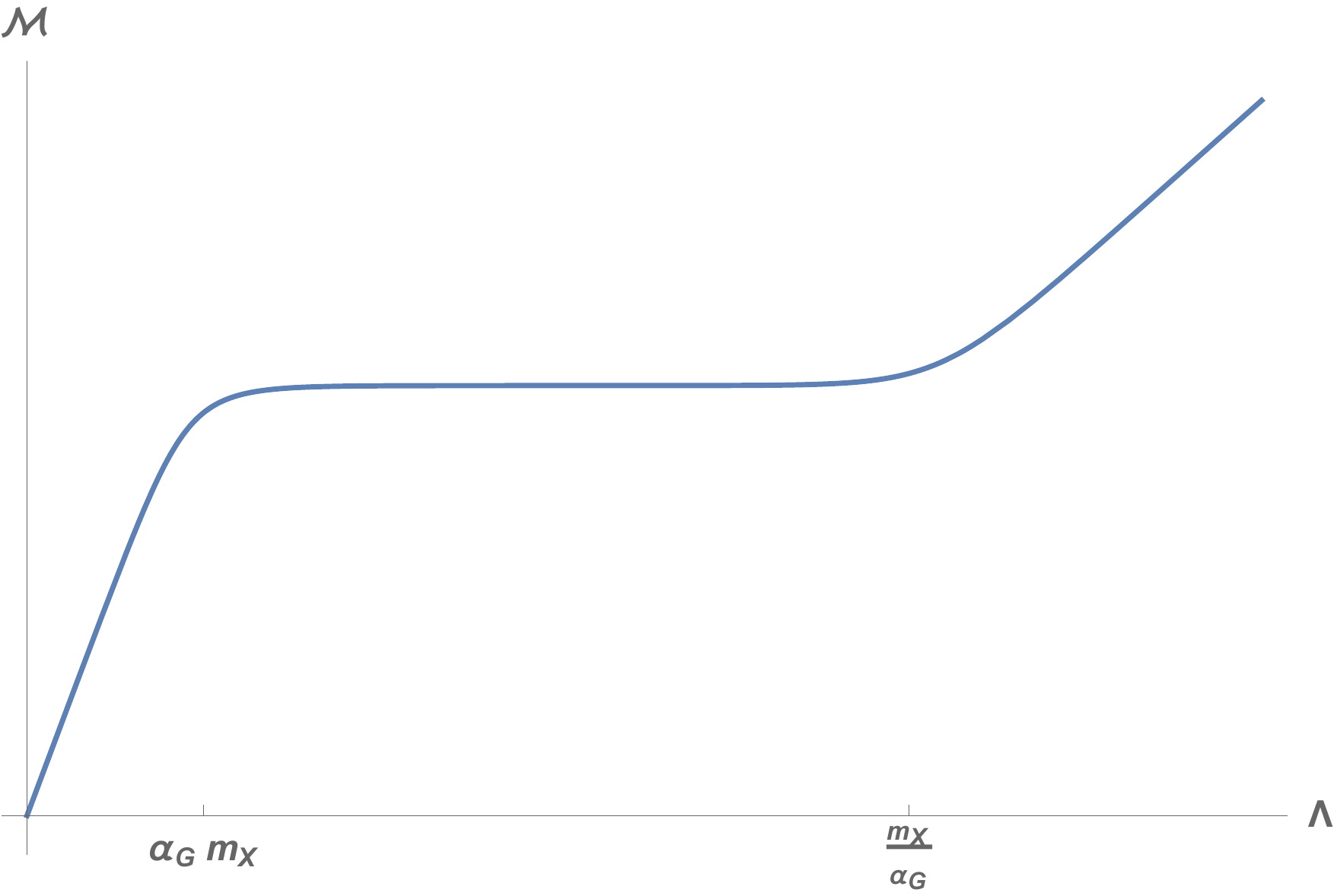}
\caption{\label{plot1} Scalar to scalar bound state amplitude as a function of the hard cut-off $\Lambda$. The amplitude converges in the non-relativistic region $\alpha_G m_X < \Lambda < m_X/\alpha_G$ before diverging in the relativistic region $\Lambda>m_X/\alpha_G$.}
\end{figure}

In fact, if we plot (\ref{M0BS}) as a function of the cut-off $\Lambda$, we can clearly see that the amplitude has a plateau for $\alpha_G m_X < \Lambda < m_X/\alpha_G$: it converges in that region before diverging for $\Lambda>m_X/\alpha_G$ when we enter the hard relativistic regime. Since the amplitude is insensitive to the value of the cut-off in the region of convergence, we can choose any value for $\Lambda$ between $\alpha_G m_X$ and $m_X/\alpha_G$ and trust the classical result. The plot is shown in Fig.\ref{plot1}.

If we take the classical amplitude cut-off at $\Lambda = m_X$ and expand it in powers of $\alpha_G$, we find
\beq
\mathcal{M}_{BS}^0 \approx i \sqrt{\frac{\pi }{2}}\alpha_G  G m_X^{5/2} \sqrt{\alpha_G  m_X}\left( 1+ \frac{\alpha_G}{3 \pi} + ... \right),
\eeq
where the first term is just (\ref{MBS}) and the next to leading order term is suppressed by a factor of $\alpha_G/3\pi$. This is consistent with our previous qualitative discussion in which we showed that the suppression factor is of order $\sim k_B^2/(\alpha_G m_X^2)=\alpha_G$. For fermion and vector SM particles the first term is exactly zero, so the first non-zero term in the cross section is suppressed by $\alpha_G^2 \sim 10^{-12}$ for a GUT scale $X$. We conclude that production of gravitational bound states without emission of external radiation is only efficient when the SM particles are scalars.

Now we can use (\ref{sigmaBS2}) to compute the total cross section for decay to SM particles. The wavefunction squared at the origin is
\beq
|\psi_0(0)|^2=\frac{\alpha_G^3 \mu^3}{\pi}=\frac{\alpha_G^3 m_X^3}{8 \pi},
\eeq
so the cross section for $S=0$ is
\beq\label{sigmaBS0}
\sigma_{BS}^0= \frac{\pi^2}{8} G^2m_X^4\alpha_G^3 \delta(s-4m_X^2)=\frac{\pi^2}{8} \left(\frac{m_X}{m_p}\right)^{10} \delta(s-4m_X^2).
\eeq
We can then plug this cross section into the Gondolo-Gelmini formula to obtain the thermally averaged cross section for production of a scalar $X$ bound state by scalar SM particles:

\beq\label{gondolo}
<\sigma_{BS}^0 v>_{SM \rightarrow X}=\frac{1}{32 T^5} \int_0^\infty ds K_1\left(\frac{\sqrt{s}}{T}\right) \sigma_{BS}^0 s^{3/2} = \frac{\pi^2}{32} \frac{m_X^3}{T^5} K_1\left(\frac{2m_X}{T}\right) \left(\frac{m_X}{m_p}\right)^{10}.
\eeq
This cross section is exponentially suppressed by the factor of $K_1(2m_X/T)$ with respect to the cross section for creating free $X$.

If the gravitational bound state can be produced from scalars, it can also decay back to scalars, with a decay rate $\Gamma_S$ that is simply related to the cross section of (\ref{sigmaBS0}) by the formula
\beq\label{sigmagamma}
\sigma_{BS}^0=\frac{8\pi^2}{m_X} \Gamma_S \delta(s-4m_X^2),
\eeq
so that the decay rate of a scalar $X$  bound state to $N_0$ scalar species is
\beq\label{gammaS}
\Gamma_S=N_0 \frac{m_X}{64} \left(\frac{m_X}{m_p}\right)^{10}\equiv N_0 \frac{\alpha_G^5 m_X}{64},
\eeq
the other decay channels being suppressed by an additional factor of $\alpha_G^2$.

The dark matter bound state cannot be created efficiently from free gravitons, since they are not in thermal equilibrium with the SM plasma, but it can decay to gravitons with a cross section given by (\ref{sigmaBS}), where $|\mathcal{M}^{S=2}_{BS} |^2 = |\mathcal{M}^{S=2}_{F} (\mathbf{0},\mathbf{0}) |^2 |\psi_0(0)|^2/m_X$ and 
\beq
\begin{split}
|\mathcal{M}^{S=2}_{F} |^2 = \frac{G^2}{2 s^2} [169 m_X^8+2 m_X^6 (53 s-58 t)+m_X^4 \left(25 s^2-42 s t+62 t^2\right) \\
+2 m_X^2 \left(8 s^3+15 s^2 t+23 s t^2-2 t^3\right) +4 s^4+10 s^3 t+11 s^2 t^2+2 s t^3+t^4]
   \end{split}
\eeq
is the amplitude squared for free $X$ production by massless spin 2 gravitons. This amplitude does not vanish for $\mathbf{k}=\mathbf{0}$, meaning that, like scalars and unlike fermions and vectors, bound state decay to gravitons is not suppressed. In fact $|\mathcal{M}^{S=2}_{F} (\mathbf{0},\mathbf{0}) |^2 = 82 G^2 m_X^4$ (compare this to the scalar case $|\mathcal{M}^{S=0}_{F} (\mathbf{0},\mathbf{0}) |^2 = 4 G^2 \pi^2 m_X^4$), and 
\beq
\sigma^{S=2}_{BS} = \frac{41}{16} G^2 m_X^4 \alpha_G^3 \delta(s-4m_X^2) = \frac{41}{16} \left(\frac{m_X}{m_p}\right)^{10} \delta(s-4m_X^2).
\eeq
Using (\ref{sigmagamma}) we obtain the decay rate to gravitons $\Gamma_G$:
\beq\label{gammaG}
\Gamma_G = \frac{41 m_X}{128 \pi^2} \left(\frac{m_X}{m_p}\right)^{10} \equiv \frac{41 \alpha_G^5 m_X}{128 \pi^2}.
\eeq
Note that $\Gamma_G/\Gamma_S = 41/(2 \pi^2 N_0) \approx 2/N_0$.

Using equation (\ref{BSamplitude}), we can also easily compute the decay rate of the first 3d excited state to scalars and gravitons. While the free amplitude in the integral is unchanged, the momentum-space wavefunction is now
\beq
\tilde{\psi}_{3d}(\mathbf{k})= \sqrt{\frac{3}{5 \pi}} \frac{2^6 3^3}{(\alpha_G \mu)^{3/2}} \frac{(\alpha_G \mu)^6 k^2}{(9 k^2+\alpha_G^2 \mu^2)^4} Y_2^m(\theta_k,\phi_k),
\eeq
where $m=-2,-1,0,1,2$ is the magnetic quantum number which specifies degenerate states with the same angular momentum. Integrating and averaging over $m$, we find that, to first order in $\alpha_G$, the decay rates are 
\begin{align}
\Gamma^S_{3d}=N_0 \frac{\alpha_G^9 m_X}{2^9 3^9 \pi}, \nonumber \\
\Gamma^G_{3d}= 41 \frac{\alpha_G^9 m_X}{2^{10} 3^9 \pi^3}.
\end{align}

\end{appendix}

\end{document}